\documentstyle[epsf,eqsecnum,aps]{revtex}
\tighten
\begin{document}
\twocolumn

\wideabs{
\title{ Electrically driven convection in a thin annular film\\
undergoing circular Couette flow}

\author{Zahir A. Daya, V.B. Deyirmenjian, and Stephen W. Morris}
\address{Department of Physics, University
of Toronto, Toronto, Ontario, Canada, M5S 1A7}
\maketitle

\begin{abstract}
We investigate the linear stability of a thin, suspended, annular film of
conducting fluid with a voltage difference applied between its inner
and outer edges.  For a sufficiently large voltage, such a film is
unstable to radially-driven electroconvection due to charges which develop
on its free surfaces.  The  film can also be
subjected to a Couette shear by rotating its inner
edge. This combination is experimentally realized using
films of smectic~A liquid crystals. In the absence of shear, the
convective flow consists of a stationary, azimuthally
one-dimensional pattern of symmetric, counter-rotating vortex pairs.
 When Couette flow is applied, an azimuthally traveling pattern
results.  When viewed in a co-rotating frame, the traveling pattern
consists of pairs of asymmetric vortices.  We calculate the neutral
stability boundary for arbitrary radius ratio $\alpha$ and Reynolds
number ${{\cal R} e}$ of the shear flow, and obtain the critical
control parameter ${\cal R}_c\,(\alpha, {{\cal R} e})$ and the
critical azimuthal mode number ${m_c\,(\alpha, {{\cal R} e})}$. The
Couette flow suppresses the onset of electroconvection, so that
${\cal R}_c\,(\alpha, {{\cal R} e}) > {\cal R}_c\,(\alpha,0)$. The
calculated suppression is compared with experiments performed at
$\alpha = 0.56 $ and $0 \leq {{\cal R} e} \leq 0.22 $.\\ \\
{\it Physics of Fluids}, {\bf 11}, 3613 (1999). See also http://mobydick.physics.utoronto.ca.
\end{abstract}

\pacs{47.20.K,47.65.+a,61.30.-v}

}
\narrowtext

\section{Introduction}
\label{intro}

We examine the effect of a circular Couette shear on radially driven convection in a two-dimensional annular fluid. Surprizingly, this remarkably ideal situation is experimentally realizeable in the electroconvection of thin, freely suspended liquid crystal films.\cite{annular98}  This system, which can be accurately described by electrohydrodynamic theory\cite{linear,gle97}, presents a unique opportunity to quantitatively study convection with a novel combination of forcing and shear. In this paper, our treatment is both theoretical and experimental; we present a complete linear stability analysis and use it to make the first quantitative comparisons with experiment.\cite{annular98}  This rather simple instablity has a rich phenomenolgy in its nonlinear behaviour, and is a promising new testing-ground for theories of bifurcations in nonlinear patterns.\cite{ch93} Since it involves radial driving forces, this system may also be interesting as an experimental analog for some geophysical instabilities.\cite{dishpan,alonso95}  

The linear instability and subsequent nonlinear evolution of flows
depend strongly on the symmetry and structure of the unstable base
flow.  One way of systematically studying this dependence is to superpose
simple flows or rotations on well-understood instabilities.  For example, the standard case of buoyancy-driven Rayleigh-B{\'e}nard convection (RBC) has been studied in the presence of rotation\cite{Chandrasekhar,goldstein9394,rotRBC,knobloch} and shearing due to an open through-flow\cite{rbc_flow,fk88}, as well as in the geophysically interesting cases of radial gravitation with rotation\cite{dishpan,alonso95}.  The phenomenology of two-dimensional (2D) electroconvection in a rectangular geometry was the subject of our previous experimental\cite{Morris,jstatphys,MorrisPRA,MDMgle,MDDMend} and theoretical\cite{linear,gle97} work. These studies made precise the degree of analogy with RBC; the electrical nature of the forcing introduces some crucial differences in detail, even at the level of linear stability.  Here we superpose this basic instability on a shear in the form of a circular Couette flow.  The annular geometry and the Couette shear naturally bring to mind another standard instability, that of Couette flow between concentric cylinders, which is unstable to three-dimensional (3D) Taylor Vortex flow (TVF).\cite{Chandrasekhar}   TVF has also been examined for the cases of an imposed axial through-flow\cite{tvf_thru} and superposed on radial electrical driving\cite{agrait88}.  Although our analysis involves a similar cylindrical geometry, the crucial difference in our case is that the flow is truly 2D, which prevents the 3D Taylor instability.\cite{drazin,wmkg95}  We will show, in fact, that the Couette shear is always stabilizing in our system.

  In summary, the unique fluid-mechanical features of our system are that it achieves the superposition of a circular Couette shear with a radially (transverse to the shear) driven convective instability in an entirely 2D flow.  This combination cannot be accurately realized in any other experimentally accessible system.  We will show that this combination leads to some interesting new phenomena.

Electroconvection results when an applied electric field acts on
space charges within a fluid, resulting in body forces which drive
flows. Electroconvection in 3D isotropic
dielectric fluids has been the subject of many previous
studies.\cite{3d_electro}  In general, the agreement between
experiment and theory has been modest, owing to difficulties in
modeling the generation of charge at the electrodes.  

 Another often-studied form of electroconvection occurs

\begin{figure}
\epsfxsize =3in
\centerline{\epsffile{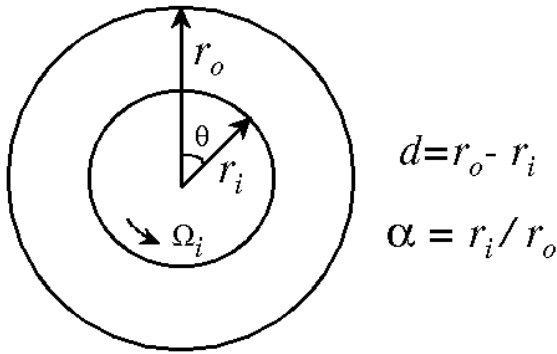}}
\vskip 0.10in
\caption{Schematic of an annular film and cylindrical coordinate
system. The film has a thickness $s \ll d$ in the $z$ direction (out of the
page), and is treated as a two-dimensional sheet.}
\label{schematic}
\end{figure}

\noindent in certain
anisotropic 3D fluids, {\it i.e.} liquid crystals, in which bulk
charge separation occurs due to a mechanism which is unique to
anisotropic fluid conductors.\cite{ecrev}  Since this mechanism does
not involve electrode effects, significant agreement between theory
and experiment has been achieved despite the complexity of the
anisotropic fluid.

In thin, suspended fluid films, another mechanism of charge
separation can take place, leading to 2D electroconvection. If a
voltage is applied across a conducting fluid film, the free surfaces
acquire a potentially unstable distribution of surface
charge.\cite{linear,gle97}  This charge must be present simply to satisfy
the standard matching conditions on electric fields across an interface.
We do not consider charge generation by electrode injection or other bulk effects.  A non-zero threshold voltage
for the onset of electroconvection results from the competition between the
electrical force on the surface charge configuration and viscous and
electrical dissipation. In a very thin, suspended film with two
free surfaces, the result is a 2D electroconvective instability,
beyond which the fluid flows in the plane of the film.

The instability we consider is in principle possible in any
isotropic, thin film of conducting fluid with a potential difference
applied across it.  To realize this experimentally, we use thin
suspended films of a liquid crystal in the smectic~A
phase.\cite{annular98,Morris,jstatphys,MorrisPRA,MDMgle,MDDMend}
These submicron films are newtonian and isotropic for flows within the plane of
the film, are immune to thickness variation, and have a very low
electrical conductivity.  The latter is required to keep the onset
voltage within an experimentally reasonable range. The fact that convection occurrs in a smectic liquid crystaline material is irrelevant except insofar as this enforces its two-dimensionality.   A similar
instability has been observed in suspended drops of other
non-smectic fluids\cite{faetti}, but these are much thicker and tend
to experience large thickness variations when flowing, and so are
much less 2D.  It is also possible to convect other, more complex,
smectic phases which are anisotropic in the film
plane.\cite{smcexp,smctheory}
We will not consider such anisotropy here.

The system we consider consists of a thin film confined in the
annular region
between concentrically placed plane circular electrodes, as shown in
Fig.~\ref{schematic}.  In the theory, both the film and the plate electrodes
lie in the $z = 0$ plane, and are treated as 2D surfaces.  A voltage
difference is applied to the electrodes which define the inner and
outer edges of the annulus.   Experimentally, we find, in the
absence of Couette shear, that the bifurcation from the quiescent
and conducting state to the convecting state results in a flow
pattern which consists of stationary, counter-rotating pairs of
vortices arranged around the circumference of the
annulus.\cite{annular98}  Each pair of vortices has reflection
symmetry and the overall pattern is spatially periodic and 1D in the
azimuthal direction.

We alter the structure and symmetry of the quiescent base state by
applying a Couette shear by rotating the inner electrode,
as shown in Fig.~\ref{schematic}. The
shear flow leads to a net mean flow in the azimuthal direction, but
unlike the previously studied cases of an imposed through-flow\cite{ch93,rbc_flow,tvf_thru}, the mean flow
here is closed on itself.  In 2D, the Couette flow by itself is
stable\cite{drazin,wmkg95}, so that shearing alone does not
introduce other instabilities.  Experimentally, we
find that the bifurcation from the Couette and conducting state to
the convecting state leads to a 1D pattern which travels
azimuthally.\cite{annular98} When this pattern is viewed in a
co-rotating frame, that is, a frame in which the pattern is stationary, we observe that it is comprised of asymmetric
counter-rotating vortex pairs.  The vortices with circulation in the
same sense as the rotation of the inner edge are narrower than
those with the opposite sense. We will show below that these features of the
nonlinear flow patterns are already present in the unstable linear modes
that we find from our stability analysis.

An important effect of the Couette flow is to suppress the onset of
electroconvection.\cite{annular98}  The Couette base state is
stabler against the radial electrical forcing than the unsheared state.  We will show that this additional stability is a feature of the shear and not of   rotation. 
The central testable result of the linear stability analysis we
present here is the degree of this suppression for various rates of
shear.  We directly compare this with our experimental results
obtained with annular smectic films and find satisfactory agreement.  In
addition, the linear stability analysis predicts the most unstable azimuthal
mode and its rate of travel around the annulus.

A crucial and somewhat subtle aspect of the phenomenon is the manner in which the electrical driving force couples with the fluid.  Although  
the fluid motion is confined to a plane, the form of the electric fields exterior to the plane, in the empty space above and
below the film, are essential.  These exterior fields are required for the self-consistent determination of the charge densities and thereby the electrical force.  Since the electric potential at any point depends on the charge density everywhere, the coupling of the two is inherently nonlocal.  This is in sharp contrast to the case of RBC, in which the coupling of the density and buoyancy body force is entirely local.\cite{linear} 

In Section~\ref{goveq}, we develop our physical model of
electroconvection in a conducting 2D fluid. This
treatment is a generalization of that presented in Ref~\cite{linear}. The
base state
solution of these electrohydrodynamic equations in an annular
geometry is presented in Section~\ref{basestate}. Fortunately, we
find that the base state configuration of potentials and charges can
be solved analytically in the annular geometry.  The electrostatic
configuration is independent of the Couette shear which forms the
base flow. The shear is imposed by applying the appropriate boundary
conditions on the flow velocity at the inner edge of the annulus.

The linear stability equations of the sheared base state for
non-axisymmetric perturbations are set up in Section~\ref{linstab}.
Unlike for the base state itself, we find that no completely
analytic solution exists of the electrostatic equations for the
perturbed quantities. In the next two sections, we present two
methods for solving the 3D,
nonlocal coupling of the potentials and charges.  In Section
\ref{local}, we make a local approximation to the electrostatic
problem which is then used to close the perturbation equations.
This yields approximate but useful analytic expressions
for all of the
field variables and the neutral stability boundary.  In
Section~\ref{nonlocal}, we relax the local approximation and use a
numerical scheme to solve the electrostatic equations.  These
results are exact, up to the accuracy of the numerical scheme.  We
find that the local approximation does very well over most of the
parameter ranges of interest.  For both solution schemes, we observe that shear
suppresses the onset of convection and drives down the value of the
azimuthal mode number of the convection pattern.

In Section~\ref{expmethod}, we review our experimental apparatus for
convecting and shearing liquid crystal films\cite{annular98}, and compare the
predictions of the previous sections with experiment.  All of the material
parameters of the fluid are known, or can be deduced from the data, so that
the theoretical results for the suppression of the onset of convection can be
compared to the
data without adjustable parameters.  We find good agreement.  In Section~\ref{discussion}, we compare our results to what is known about other standard instabilities, particularly  RBC, under shear and/or rotation.  
Section~\ref{conclusion} is a brief summary and conclusion.

\section{The Governing Equations}
\label{goveq}

In this section, our physical model describing electroconvection in
a suspended thin film is presented.  Our analysis here is similar
to that of Ref~\cite{linear}.  Unless
otherwise stated, all operators, field variables, and material
parameters are 2D quantities.

The film is treated as a 2D conducting fluid in the $z=0$ plane,
with areal
density $\rho$, molecular viscosity $\eta$, and conductivity $\sigma$. We
consider incompressible fluids, so that the
2D velocity field, ${\bf u}$, is divergence free,
\begin{equation}
\nabla \cdot {\bf u} = 0 \,. \label{incomp}
\end{equation}
The Navier-Stokes equation with an electrical body force,

\begin{equation}
\rho \biggl[\frac{\partial {\bf u}}{\partial t} + ({\bf u} \cdot
\nabla) {\bf
u}\biggr]  =  -\nabla P + \eta {\nabla} ^{2} {\bf u} + q {\bf E} \,,
\label{NS}
\end{equation}
governs the fluid flow, where $\nabla$, $P$, $q$, and ${\bf E}$ are
the 2D
gradient operator, pressure field, surface charge density, and
electric field in the film plane, respectively. The term $q{\bf E}$
is the
electric force acting on the surface charge density. The charge
continuity
equation
\begin{equation}
\frac{\partial q}{\partial t} =
-\nabla \cdot ( q{\bf u} + \sigma {{\bf E}}) \,,
\label{CC}
\end{equation}
takes into account the convective and conductive current densities,
$q{\bf u}$
and  $\sigma {\bf E}$ respectively.  Diffusion of charge in the plane of the
film is neglected.

Subscript three will be used to denote three-dimensional (3D) differential
operators, material parameters and field variables.  The 3D
electric potential $\Psi_3$ is governed by the 3D Laplace equation,
\begin{equation}
\nabla_3^2 \Psi_3 = 0 \,,
\label{LAP}
\end{equation}
where $\nabla_3$ is the 3D gradient
operator.  The coupling of $\Psi_3$ with the 2D charge density $q$ is
specified by requiring $\Psi_3$ to satisfy certain boundary conditions on the
$z=0$ plane.    The
surface charge density $q$ depends on the discontinuity in the
$z-$derivative of
$\Psi_3$ across the two surfaces of the film:

\begin{eqnarray}
q & = & - {\epsilon_0} \frac{\partial \Psi_3}{\partial z}{\Bigg|_{z=0^+}}
 + {\epsilon_0} \frac{\partial \Psi_3}{\partial z}{\Bigg|_{z=0^-}}
   \,, \nonumber \\
& = & -2 {\epsilon_0} \frac{\partial \Psi_3}{\partial z}{\Bigg|_{z=0^+}},
\label{qdef}
\end{eqnarray}
where $\epsilon_0$ is the permitivity of free space.  If $q$ is known,
Eqs.~\ref{qdef} constitute Neumann conditions on $\Psi_3$ on the film,
while Dirichlet conditions, described below, hold on the electrodes.  If
instead Dirichlet conditions are specified on the film, Eqs.~\ref{qdef}
can be used to determine $q$.

  The 2D and 3D
potentials are related via $\Psi=\Psi_3|_{z=0}$.  Equations~\ref{qdef}
relate the surface charge density to the discontinuity in the $z$ component
of the 3D electric field ${\bf E}_3= - \nabla_3 \Psi_3$ across the film
plane.  On the other hand, the $x$ and $y$ components of ${\bf E}_3$ which
form the 2D electric field ${\bf E}= - \nabla \Psi$, are continuous across
the film.  This continuity is required by the usual matching conditions for
electric fields across the surfaces of dielectrics.  Note that it is the 2D
quantity ${\bf E}$ and not ${\bf E}_3$ that appears in equations~\ref{NS} and
\ref{CC}. One cannot simply use a Maxwell equation to eliminate the charge
density in favor of the field\cite{smcfootnote} because the 2D quantities in question are confined to a plane embedded in a 3D, otherwise empty, space and in general $\nabla \cdot {\bf E} \neq
q/\epsilon_0$.  

Equations~\ref{incomp}-~\ref{qdef}, together with the appropriate
boundary
conditions on the electrodes, model our system.  We consider the
electrohydrodynamic limit where  magnetic fields and the resultant Lorentz
forces are negligible.  One can also show that dielectric polarization
effects are negligible in the limit of a thin film.\cite{linear}

In our subsequent analysis, we define the stream function $\phi$ by
\begin{equation}
{\bf u} = \nabla \times {\vec \phi} \,, \label{stream function}
\end{equation}
where ${\vec \phi} = {\phi} {\bf \hat z}$. Using
Eq.~\ref{stream function},
${\bf E}= - \nabla \Psi$ and
eliminating the pressure field by applying the curl operator,
Eqs.~\ref{NS} and~\ref{CC} reduce to
\begin{eqnarray}
\rho \biggl[\frac{\partial}{\partial t}
+ ( \nabla \times {\vec \phi} ) \cdot \nabla \biggr]
( \nabla \times \nabla \times {\vec \phi} ) \nonumber \\
- \eta \nabla^2 ( \nabla \times \nabla \times {\vec \phi} )
+ ( \nabla q \times \nabla \Psi ) & = & 0 \,, \label{NS2}
\end{eqnarray}
and
\begin{equation}
\frac{\partial q}{\partial t}  +  ({\nabla \times
{\vec \phi})\cdot \nabla} q - \sigma \nabla^2 \Psi = 0 \,. \label{CC2}
\end{equation}

We render these equations dimensionless by rescaling the length, time,
and electric
potential by $d$, $\epsilon_{0}d/\sigma$, and $V$, respectively,
where $d$ and
$V$ are the cross-film width and potential difference. It follows
that the
stream function and charge density are
nondimensionalized by $\sigma d/\epsilon_0$ and $\epsilon_0 V/d$.
Applying this rescaling to  Eqs.~\ref{NS2}, \ref{CC2},
\ref{LAP}, and \ref{qdef} gives
\begin{eqnarray}
\Biggl[ \nabla^2 - \frac{1}{{\cal P}}\frac{\partial}{\partial t}\Biggr]
(\nabla \times \nabla \times {\vec \phi}) + {\cal R}
\biggl(\nabla\Psi \times \nabla q \biggr) = \nonumber \\
\frac{1}{{\cal P}} \biggl( (\nabla \times {\vec \phi}) \cdot
 \nabla \biggr) ( \nabla \times \nabla \times {\vec \phi}) \,,
\label{NDNS}
\end{eqnarray}
\begin{equation}
\frac{\partial q}{\partial t} + ( \nabla \times {\vec \phi} )
\cdot \nabla q - \nabla^2 \Psi = 0 \,, \label{NDCC}
\end{equation}
\begin{eqnarray}
\nabla_3^2 \Psi_3 &=& 0 \,, \label{NDLAP} \\
q &=& -2 \frac{\partial \Psi_3}{\partial z}{\Biggl|}_{z=0^+} \,,
\label{NDqdef}
\end{eqnarray}
where, the dimensionless parameters
\begin{equation}
{\cal R} \equiv \frac{\epsilon_0^2 V^2}{\sigma \eta}=
\frac{\epsilon_0^2 V^2}{\sigma_3 \eta_3 s^2} \hspace{5mm} {\rm and}
\hspace{5mm}
{\cal P} \equiv
\frac{\epsilon_0 \eta}{\rho \sigma d}=\frac{\epsilon_0
\eta_3}{\rho_3 \sigma_3
s d} \,, \label{Rayleigh}
\end{equation}
are analogous to the Rayleigh and Prandtl numbers in the
Rayleigh-B\'enard
problem. Here, $s$ is the thickness of the film and our 2D treatment assumes
that $s \ll d$.  The 2D material parameters are related to their
three-dimensional counterparts by $\sigma = \sigma_3
s$, $\eta = \eta_3 s$, and $\rho = \rho_3 s$.  The control parameter
${\cal R}$
is proportional to the square of the applied voltage difference, but
independent of the film width $d$.
${\cal P}$ is the ratio of the charge
relaxation time scale in a film $\epsilon_0 d / \sigma_3 s$ to the
viscous relaxation time scale $\rho_3 d^2 / \eta_3$.

Equations
\ref{NDNS}-\ref{NDqdef}, together with appropriate boundary conditions,
describe  electroconvection in a thin conducting film suspended in otherwise
empty space, for any 2D arrangement of the film and electrodes.

\section{The base state}
\label{basestate}

We now specify the governing equations introduced in
Section~\ref{goveq} to the case of an annular film, and solve them for the
case of a general Couette shear flow which forms the potentially unstable
base state.

We employ cylindrical
coordinates $(r,\theta,z)$ as in
Fig.~\ref{schematic}. The film is suspended between two circular
electrodes which cover the remainder of the $z=0$ plane. The inner electrode
has a radius $r_i$ and is at potential
1 in our
dimensionless units. The outer
electrode, which occupies the $z=0$ plane for $r > r_o$, is at zero
potential.   The
cross-film width is $r_o - r_i = 1$ in dimensionless units and we define the
radius ratio,
\begin{equation}
\alpha = r_i/r_o,
\end{equation}
so that
\begin{equation}
r_i = \frac{\alpha}{1-\alpha}\,,\hspace{1cm} r_o =
\frac{1}{1-\alpha}\,.\hspace{1cm}
\end{equation}

Rotation of the inner electrode about the central $r=0$ axis
produces a Couette shear in the base state.  We will denote base state
variables by the superscript zero. Under shear, the radial
derivative of base state stream function is given by
\begin{equation}
\partial_r\phi^{(0)}(r) = \frac{\alpha^2 \Omega}{1 - \alpha^2}
\biggl( r -
\frac{1}{r(1-\alpha)^2}\biggr) \,, \label{phizero}
\end{equation}
where $\Omega$ is the dimensionless angular
rotation frequency of the inner electrode.  If the fluid is not sheared, the
base state velocity field is zero and $\phi^{(0)}(r) \equiv 0$.  The
strength of the shear is described by a Reynolds number, with the velocity determined by the motion of the inner edge and the length by the film width,
\begin{equation}
{{\cal R}e}  = {r_i  \Omega  \over {\cal P}} \,, \label{reynolds}
\end{equation}
where $\cal P$ is the Prandtl number given by Equation~\ref{Rayleigh}.

There is no loss of generality by treating only rotations of the inner
electrode.  Since our system is 2D, one can always reduce independent rotations of both electrodes to
this case by transforming to a rotating reference frame in which the outer
edge is stationary.  The transformation introduces a coriolis term into
Eq.~\ref{NS}, which can simply be absorbed into the pressure.\cite{alonso95}  As a result there are important differences in the stability of 2D and 3D systems, and in particular for rigid rotation. We will return to this point in section \ref{discussion} below.

The base state potential $\Psi_3^{(0)}(r,z)$ and charge density
$q^{(0)}(r)$ are independent of the base state shear flow.  They are
determined by the electrostatic boundary value problem given by
Eqs.~\ref{NDLAP} and \ref{NDqdef}, with Dirichlet boundary
conditions on the $z=0$ plane
\begin{equation}
\Psi^{(0)} (r) = \Psi^{(0)}_3(r,0) = \nonumber \\
\left\{
\begin{array}{cc}
   1  & 0 \leq r \leq r_i  \\
  \frac{1}{ln(\alpha)}[ln(1-\alpha)+ln(r)]  & r_i \leq r \leq
r_o \\
  0  & r \geq r_o\,. \\

\end{array}
\right.
\end{equation}
The boundary condition for $r_i \leq r \leq r_o$ is found by treating the
annular film as a 2D ohmic conductor subject to a dimensionless potential of
$1$ at the inner edge and $0$ at the outer, and requiring the continuity of
the 2D current density. The logarithmic form follows from the cylindrical
geometry.

\begin{figure}
\epsfxsize =3.2in
\centerline{\epsffile{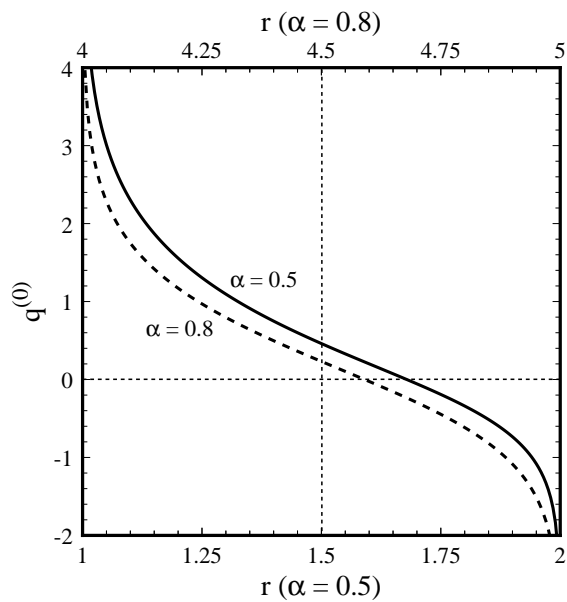}}
\vskip 0.1in
\caption{Base state charge density $q^{(0)}(r)$ at $\alpha = 0.5$
and $\alpha = 0.8$. $q^{(0)}$ diverges at the edges of the film. }
\label{q0_plot}
\end{figure}

The Laplace equation~\ref{NDLAP} for the potential in the half-space $z  > 0$
is
solved by the ansatz
\begin{equation}
\Psi_3^{(0)}(r,z) =  {\int_0}^\infty dk ~ e^{-kz} J_0 (kr) A(k) \,,
\end{equation}
where $J_0$ is the zeroth order Bessel function. Inversion of the
above equation results in
\begin{equation}
A(k) = k {\int_0}^\infty dr~ r \Psi_3^{(0)}(r,0) J_0 (kr) \,.
\end{equation}
Hence, the base state surface charge density is given by
\begin{equation}
q^{(0)}(r) = 2{\int_0}^\infty dk~ k^2 J_0 (kr) \Biggl[ {\int_0}^\infty
d\zeta ~\zeta \Psi_3^{(0)}(\zeta,0) J_0 (k\zeta)\Biggr],
\end{equation}
where ${\zeta}$ is a dummy integration variable. Evaluating the
integrals\cite{PBM,GR}, we find
\begin{equation}
q^{(0)} (r) = \frac{2}{\ln{\alpha}} \Biggl[ \frac{1}{r} F\biggl(
\frac{1}{2},  \frac{1}{2};1; \frac{{r_o}^2}{r^2}\biggr) -
\frac{1}{r_i}F\biggl(\frac{1}{2},  \frac{1}{2};1;
\frac{r^2}{{r_i}^2}\biggr)\Biggr],\label{qhyper}
\end{equation}
where $F$ is the hypergeometric function $_2F_1$. This function is plotted
in Fig.~\ref{q0_plot} for two values of $\alpha$.  As $\alpha \rightarrow 1$,
$q^{(0)}$ approaches the base state charge density for a laterally unbounded
rectangular film, which is odd-symmetric about the midline of the film. This
symmetry helped simplify the analysis in that case.\cite{linear}
However, for $0 < \alpha < 1$, the annular base state charge density
$q^{(0)}$ is neither even nor odd about the midline, so the analysis here is
more complicated.  This deviation from odd symmetry
is larger for smaller $\alpha$.

The surface charge density shown in Fig.~\ref{q0_plot} is ``inverted'' in
the sense that the positive charges lie close to the positive, inner
electrode and
negative charges are near the outer electrode. This unstable
surface charge configuration gives rise to an electroconvective instability,
much like the unstably stratified density configuration which produces
Rayleigh-B{\'e}nard convection.\cite{linear}  The divergences of $q^{(0)}$ at
the edges of the film are a consequence of the idealized
geometry in which the electrodes have zero thickness.  
\vskip 0.2in

This idealization
leads to the boundary conditions in Eq.~(3.5)
, whose radial
derivative is discontinuous at $r = r_i$ and $ r =r_o$.  This is reflected in
Eq.~\ref{qhyper} where ${F\Bigl(
\frac{1}{2},  \frac{1}{2};1; 1\Bigr)}$ is indeterminate, so that $q^{(0)}$
diverges at the edges of the film.

\section{Linear Stability Analysis}
\label{linstab}
In this section, we test the stability of the axisymmetric base state to
non-axisymmetric perturbations.  The perturbed quantities will be denoted by
the superscript one, and are defined by

\begin{eqnarray}
\nonumber \phi(r,\theta) & = & \phi^{(0)}(r) + \phi^{(1)}(r,\theta)
\,, \\
\nonumber q(r,\theta) & = & q^{(0)}(r) + q^{(1)}(r,\theta) \,, \\
\Psi(r,\theta) & = & \Psi^{(0)}(r) + \Psi^{(1)}(r,\theta) \,, \\
\nonumber \Psi_3(r,\theta,z) & = & \Psi_3^{(0)}(r,z) +
\Psi_3^{(1)}(r,\theta,z)
\,.
\label{perturbs}
\end{eqnarray}
Substitution of the perturbed field variables into
Eqs.~\ref{NDNS}-\ref{NDqdef} and retaining only the terms which are
linear in the perturbed quantities yields

\begin{eqnarray}
\Biggl[ \nabla^2 - \frac{1}{{\cal P}}\Biggl( \frac{\partial}{\partial t}
 - \frac{1}{r} \frac{\partial \phi^{(0)}}{\partial r}
   \frac{\partial}{\partial \theta} \Biggr) \Biggr] (\nabla^2 \phi^{(1)})
\nonumber \\
- \frac{{\cal R}}{r} \Biggl( \frac{\partial \Psi^{(0)}}{\partial r}
  \frac{\partial q^{(1)}}{\partial \theta}
-\frac{\partial \Psi^{(1)}}{\partial \theta}
 \frac{\partial q^{(0)}}{\partial r}\Biggr) = \nonumber \\
 \frac{1}{r {\cal P}} \frac{\partial \phi^{(1)}}{\partial \theta}
 \frac{\partial}{\partial r}
 \biggl( \frac{\partial^2}{\partial r^2}
+\frac{1}{r}\frac{\partial}{\partial r} \biggr)\phi^{(0)}  \,,
\label{NDNS_lin} \\
\frac{\partial q^{(1)}}{\partial t} + \frac{1}{r} \Biggl(
 \frac{\partial \phi^{(1)}}{\partial \theta}
 \frac{\partial q^{(0)}}{\partial r}
-\frac{\partial q^{(1)}}{\partial \theta}
 \frac{\partial \phi^{(0)}}{\partial r} \Biggr) \nonumber \\
- \nabla^2 \Psi^{(1)}  = 0 \,, \label{NDCC_lin} \\
\nabla_3^2 \Psi_3^{(1)} = 0 \,, \label{NDLAP_lin} \\
q^{(1)} = -2 \frac{\partial \Psi_3^{(1)}}{\partial z}
  {\Biggl|}_{z=0^+} \,, \label{NDqdef_lin}
\end{eqnarray}
where
\begin{eqnarray}
\nonumber \nabla^2  \equiv  \frac{\partial^2}{\partial r^2}
+\frac{1}{r}\frac{\partial}{\partial r}
+\frac{1}{r^2}\frac{\partial^2}{\partial \theta^2} \hspace{0.5cm} {\rm and}
\hspace{0.5cm} \nabla_3^2 \equiv  \nabla^2 +
\frac{\partial^2}{\partial z^2} \,.
\end{eqnarray}
The variables $\phi^{(1)}$, $\Psi^{(1)}$, and $\Psi^{(1)}_3$ satisfy the
following boundary conditions:

\begin{eqnarray}
\phi^{(1)}(r_i,\theta)=\partial_r \phi^{(1)}(r_i,\theta) = \nonumber \\
\phi^{(1)}(r_o,\theta)=\partial_r \phi^{(1)}(r_o,\theta) =  0 \,,
\label{directbc1} \\
\Psi^{(1)}(r_i,\theta) = \Psi^{(1)}(r_o,\theta) = 0 \,,
\label{directbc2} \\
\Psi_3^{(1)}(r,\theta,z) \rightarrow 0~ &\text{for}&~ z \rightarrow
\pm \infty
 \,. \label{directbc3}
\end{eqnarray}
Equation \ref{directbc1} is a consequence of rigid boundary
conditions on the
fluid flow.  The Dirichlet boundary conditions for the perturbed potential
$\Psi_3^{(1)}$ on the $z=0$ plane are
\begin{equation}
\Psi_3^{(1)}(r,\theta,0) = \left\{
\begin{array}{ccc}
   0 &\text{for} & 0 \leq r \leq r_i  \\
  \Psi^{(1)}(r,\theta) & \text{for} & r_i \leq r \leq r_o \\
  0 & \text{for} & r \geq r_o~. \\
\end{array}
\right.  \label{directbc4}
\end{equation}

We decompose the perturbations into products of the following form,
\begin{equation}
 \left(
\begin{array}{c}
\phi^{(1)}(r,\theta)  \\
q^{(1)}(r,\theta)  \\
\Psi^{(1)}(r,\theta)  \\
\Psi_3^{(1)}(r,\theta,z)  \\
\end{array}
\right) = \left(
\begin{array}{c}
{\phi}_m(r)  \\
{q}_m(r)  \\
{\Psi}_m(r)  \\
{\Psi}_{3m}(r,z)  \\
\end{array}
\right) e^{i m\theta + \gamma t}  \,,
\label{normalmode}
\end{equation}
where the azimuthal mode number $m$ is an integer which corresponds to the
number of vortex pairs in the pattern. The functions
$\phi_m$, $\Psi_m$, and $\Psi_{3m}$ satisfy the same boundary
conditions as the perturbations, Eqs.~\ref{directbc1}
-\ref{directbc4}. The growth rate $\gamma$ may be complex.

Substitution of Eq.~\ref{normalmode} into
Eqs.~\ref{NDNS_lin}-\ref{NDqdef_lin} gives
\begin{eqnarray}
\biggl(D_*D - \frac{m^2}{r^2}\biggr)^2 {\phi}_m  \nonumber \\
 -\frac{1}{{\cal P}} \biggl( \gamma-\frac{imD\phi^{(0)}}{r} \biggr)
  \biggl(D_*D - \frac{m^2}{r^2}\biggr) {\phi}_m   \nonumber \\
- \frac{im{\cal R}}{r} \biggl( (D\Psi^{(0)}){q}_m
      - (Dq^{(0)}){\Psi}_m   \biggr)
  =  \nonumber \\
\frac{im{\phi}_m}{r{\cal P}}D(D_*D{\phi}^{(0)}) \,,  \label{NDNS_mode}
\end{eqnarray}
\begin{eqnarray}
\biggl(D_*D - \frac{m^2}{r^2} \biggr) {\Psi}_m 
 - \biggl( \frac{imDq^{(0)}}{r} \biggr){\phi}_m \nonumber \\
 - \biggl( \gamma -  \frac{imD\phi^{(0)}}{r}\biggr) {q}_m
  =  0  \,, \label{NDCC_mode}
\end{eqnarray}
\begin{eqnarray}
\biggl( D_*D - \frac{m^2}{r^2} + \frac{\partial^2}{\partial z^2} \biggl)
  {\Psi}_{3m}  =  0 \,, \label{NDLAP_mode} 
\end{eqnarray}
\begin{eqnarray}
{q}_m  =  -2 \frac{\partial {\Psi}_{3m}}{\partial z}{\Biggl|}_{z=0^+}
\,, \label{NDqdef_mode}
\end{eqnarray}
where $D \equiv \partial_r$ and $D_* \equiv D + 1/r$. Henceforth, we limit
our discussion to a base state flow which is either quiescent or Couette. In
these cases $D(D_*D{\phi}^{(0)}) \equiv 0$ and the right hand side of
Eq.~\ref{NDNS_mode} is identically zero.
\cite{drazin}  It can be shown that in the limit $\alpha \rightarrow 1$, and
for zero shear ${\phi^{(0)}} \equiv 0$,
Eqs.~\ref{NDNS_mode}-\ref{NDqdef_mode} reduce to the linear stability
equations for electroconvection in a laterally unbounded strip.\cite{linear}

We write the complex growth exponent $\gamma$ as $\gamma^r + i\gamma^i$.  In
order to find the conditions for marginal stability, we set the real part
$\gamma^r = 0$. Our task is then to solve
Eqs.~\ref{NDNS_mode}-\ref{NDqdef_mode} for a given $\alpha$, ${\cal P}$ and
${\cal R}e$, by determining consistent values of ${\cal R}$ and $\gamma^i$ for
each
$m$. The rate of azimuthal travel of each marginally stable mode around the
annulus is $\gamma^i/m$. The marginal stability boundary, which is defined
only at discrete $m$, has a minimum at the critical values $m_c$, ${\cal
R}_c$, while the critical mode travels at $\gamma_c^i/m_c$.

We employ the following basic expansions:

\begin{eqnarray}
{\phi}_m(r) & = & \sum_{n} A_n
       {\phi}_{m;n}(r) \,, \label{phi_m} \\
{\Psi}_m(r) & = &
      \sum_{n} A_n {\Psi}_{m;n}(r) \,, \label{Psi_m} \\
{\Psi}_{3m}(r,z) & = &
      \sum_{n} A_n {\Psi}_{3m;n}(r,z) \,, \label{Psi_3m} \\
{q}_m(r) & = & \sum_{n} A_n
       Q_{m;n}(r) \,, \label{q_m}
\end{eqnarray}
where the $A_n$ are amplitudes. The expansion eigenfunctions
${\phi}_{m;n}(r)$, ${\Psi}_{m;n}(r)$, and ${\Psi}_{3m;n}(r,z)$ satisfy the
following boundary conditions:

\begin{eqnarray}
{\phi}_{m;n}(r_i)=\partial_r {\phi}_{m;n}(r_i) = \nonumber \\
{\phi}_{m;n}(r_o)=\partial_r {\phi}_{m;n}(r_o) = 0\,,
\label{directbc1mn} \\
{\Psi}_{m;n}(r_i) = {\Psi}_{m;n}(r_o) = 0\,,
\label{directbc2mn} \\
{\Psi}_{3m;n}(r,z) \rightarrow 0 ~ &\text{for}&~ z \rightarrow \pm
\infty
 \,, \label{directbc3mn}
\end{eqnarray}
\begin{equation}
{\Psi}_{3m;n}(r,z = 0) = \left\{
\begin{array}{ccc}
   0 & \text{for} & 0 \leq r \leq r_i  \\
  {\Psi}_{m;n}(r) & \text{for} & r_i \leq r \leq r_o \\
  0 & \text{for} & r \geq r_o \\
\end{array}
\right. \,. \label{directbc4mn}
\end{equation}
The functions ${\phi}_{m;n}(r)$, which satisfy the `rigid' boundary
conditions given in Eq.~\ref{directbc1mn}, can be identified with the
Chandrasekhar cylinder functions\cite{Chandrasekhar},
\begin{eqnarray}
{\phi}_{m;n}(r) = {\cal C}_{m;n}(r) = J_m(\beta_{mn} r) 
+B_{mn} Y_m(\beta_{mn} r) \nonumber \\
+ C_{mn} I_m(\beta_{mn} r) +D_{mn} K_m(\beta_{mn}
r) \,.
\label{Cmn}
\end{eqnarray}
The boundary conditions Eq.~\ref{directbc2mn}, imply that the 2D potential
expansion function ${\Psi}_{m;n}$ can be further expanded in a series of
functions of the form,
\begin{equation}
\psi_{m;p}(r) = J_m(\chi_{mp} r) + b_{mp} Y_m(\chi_{mp} r) \,.
\label{psimn}
\end{equation} The functions ${\cal C}_{m;n}$ and $\psi_{m;p}$, along with
their associated constants $\beta_{mn}$, $B_{mn}$, $C_{mn}$, $D_{mn}$,
$\chi_{mp}$, and $b_{mp}$ are described in detail in the Appendix.

In the next two sections, we use these expansions to solve
Eqs.~\ref{NDNS_mode}-\ref{NDqdef_mode}, first approximately and then in
general.

\section{Linear Stability: Local approximation}
\label{local}

The main barrier to solving Eqs.~\ref{NDNS_mode}-\ref{NDqdef_mode} lies in
the difficult nonlocal coupling between ${\Psi}_m$ and ${q}_m$ in
Eqs.~\ref{NDNS_mode} and \ref{NDCC_mode} which is required by the 3D
electrostatic equations \ref{NDLAP_mode} and \ref{NDqdef_mode}. In this
section, we circumvent this problem by making an approximation.  This yields
expressions which give some insight into the general linear stability
problem.

To make the approximation, we replace the 3D Eqs.~\ref{NDLAP_mode} and
{}~\ref{NDqdef_mode} with the following simple 2D closure relation:
\begin{equation}
\label{APPROXCL}  \Biggl(D_{*}D - \frac{m^2}{r^2} \Biggr)
{f}_m{\Psi}_m = -{q}_m.
\end{equation}
In the above expression, ${f}_m$ is a closure factor which is independent of
$r$, and is to be specified. As we show below, a consequence of this
approximation is that the charge density and the 2D potential are related
pointwise, or locally.

This approximation is motivated by following physical
reasoning.  If instead of an annular film, one considers an
annular column, with a height much larger than its width, then there is a
straightforward Poisson relation between a {\em bulk} charge density and the
3D potential inside the column.  If the 3D potential is independent of $z$,
and is equal to ${\Psi}_m(r) e^{im\theta}$, there are no free surfaces to
consider and one has in place of Eqs.~\ref{NDLAP_mode} and \ref{NDqdef_mode}
the relation ${ \Bigl(D_{*}D - {m^2}/{r^2} \Bigr){\Psi}_m = -{q}_m}$.  If one
now hypothesizes that the charge density retains its radial profile when the
bulk is `squeezed down' to a 2D film, then one must include only an $r$
independent
scaling factor, $f_m$, as in Eq.~\ref{APPROXCL}.

We make use of \ref{Psi_m} and further expand each ${\Psi}_{m;n}$ using
Eq.~\ref{psimn} so that
\begin{equation}
{\Psi}_m =\sum_n A_n {\Psi}_{m;n} = \sum_n A_n
\sum_p B_{m;np} \psi_{m;p} \,,
\label{EXPWF1}
\end{equation}
where the $B_{m;np}$ are constants. It then follows from
Eqs.~\ref{APPROXCL} and Eq.~\ref{eigeneq3} that
\begin{eqnarray}
{{q}_m}= \sum_n A_n \sum_p B_{m;np} f_{m;p} {\chi_{mp}}^2
\psi_{m;p} = \nonumber \\
 \sum_n A_n \sum_p
B_{m;np}  q_{m;p}\,,
 \label{EXPWF2}
\end{eqnarray}

\begin{figure}
\epsfxsize =3in
\centerline{\epsffile{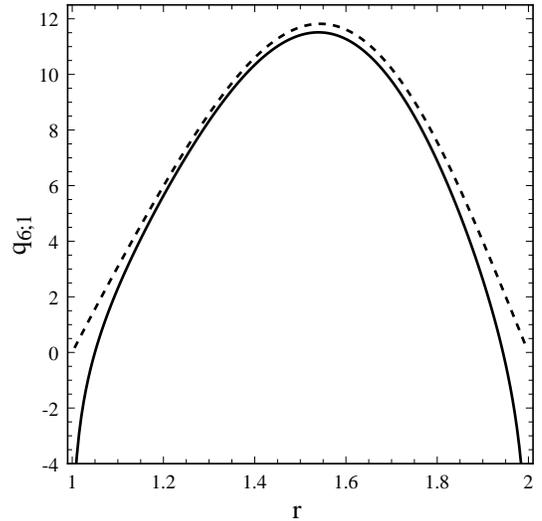}}
\caption{The charge density perturbation $q_{6;1}(r)$ at $\alpha = 0.50$,
as determined by the local approximation of Section~\protect\ref{local}
(dashed line) and the nonlocal solution of Section~\protect\ref{nonlocal}
(solid line). The nonlocal result diverges to $-\infty$ at the edges of the
film.}
\label{qcomparison}
\end{figure}

\noindent where
\begin{equation}
\label{QWF} q_{m;p} = f_{m;p} {\chi_{mp}}^2 \psi_{m;p}\,,
\end{equation}
which demonstrates the pointwise, local relation between the potential
on the film and its surface charge density in this approximation.

A choice for $f_{m;p}$ can be made by considering
the empty upper half space $z>0$ with homogeneous
boundary conditions at infinity and Dirichlet boundary conditions such that
the potential is equal to $\psi_{m;p}(r) e^{i m \theta}$ everywhere on the $z
= 0$ plane.  This boundary condition is smooth, unlike the piecewise smooth
conditions in Eq.~\ref{directbc4mn}, which respect the sharp edges of the
annulus. Then the 3D potential for $z \geq 0$ satisfies the equation,
\begin{equation}
\Biggl(D_{*}D - \frac{m^2}{r^2}  + \frac{\partial^2}{\partial z^2}\Biggr)
\psi_{m;p} e^{-k_{mp} z} ~= 0 \,.
\end{equation}
Using the eigenvalue relation Eq.~\ref{eigeneq3}, it follows that
$k_{mp} = \chi_{mp}$.  The corresponding surface charge density is thus
\begin{equation}
\label{SCDQMN} q_{m;p} = -2 \frac{\partial }{\partial z} \psi_{m;p}
e^{-\chi_{mp}
z}{\Biggl|}_{z=0^+} = 2 \chi_{mp} \psi_{m;p}.
\end{equation}
Comparing Eq.~\ref{SCDQMN} with Eq.~\ref{QWF} leads to the following choice
for the closure factor:
\begin{equation}
\label{f} f_{m;p} = \frac{2}{\chi_{mp}} \;.
\end{equation}
For the remainder of this section, we adopt this method of closing the
equations.  Fig.~\ref{qcomparison} shows a plot of the approximate charge
density, $q_{6;1}$ at $\alpha = 0.5$, corresponding to a potential
$\psi_{6;1}$ on the film.  It is compared to a more accurate numerical
solution discussed in Section~\ref{nonlocal}, below.  As might be expected,
the approximation
is accurate except close to the edges of the film.

We can now solve the remaining 2D equations using the expansions in
Eqs.~\ref{phi_m},~\ref{Cmn},~\ref{EXPWF1} and \ref{EXPWF2}.  We consider the
simplest case: a single expansion mode, so that $A_1 = 1$ and $A_n = 0$ for
$n  > 1$. Similarly, we truncate the expansions of the potential and charge
density at $p = 1$ so that $B_{m;1p} \equiv 0$ for $p > 1$.  The expansion
coefficient
$B_{m} \equiv B_{m;11}$ is complex.  Thus, using Eq.~\ref{f},
Eqs.~\ref{phi_m}, \ref{EXPWF1} and \ref{EXPWF2} reduce to
\begin{eqnarray}
\nonumber {{\phi}_m} &=&  {\cal C}_{m;1}, \\
 \label{APPROXEX} {\Psi}_m &=&  (B^r_m + i B^i_m)
\psi_{m;p}, \\
\nonumber {{q}_m} &=& 2  (B^r_m + i B^i_m) {\chi_{m1}} \psi_{m;1}.
\end{eqnarray}
Substitution of Eq.~\ref{APPROXEX} into Eq.~\ref{NDCC_mode}, with $\gamma =
i \gamma^i$ results
in an equation, the real and imaginary parts of which are
\begin{equation}
\label{APPROXCCR}  B^r_m {\chi_{m1}} \psi_{m;1} = 2 \biggl(
\gamma^i -  \frac{m D{\phi}^{(0)}}{r}\biggr)  B^i_m
 \psi_{m;1}\,,
\end{equation}
\begin{eqnarray}
\label{APPROXCCI}  B^i_m {\chi_{m1}}^2 \psi_{m;1}+
2 \biggl( \gamma^i
-  \frac{m D{\phi}^{(0)}}{r}\biggr)  B^r_m
{\chi_{m1}}
\psi_{m;1} \nonumber \\
+ \biggl( \frac{mD{q}^{(0)} }{r} \biggr){\cal C}_{m;1} = 0\,,
\end{eqnarray}
respectively. Eliminating $B^r_m$ from the above pair of equations results in
\begin{eqnarray}
\label{APPROXCCI2}  \Biggl[1 + \frac{4}{{\chi_{m1}}^2} \biggl( \gamma^i -
\frac{m
D{\phi}^{(0)}}{r}\biggr)^2  \Biggr] B^i_m
{\chi_{m1}}^2
\psi_{m;1} \nonumber \\
+ \biggl( \frac{mD{q}^{(0)} }{r} \biggr){\cal C}_{m;1} = 0.
\end{eqnarray}
Multiplying by $\psi_{m;1}$, and integrating to form inner products,
\begin{eqnarray}
\bigl< ~...~\bigr > & = & \int_{r_i}^{r_o} \, ~...~ r dr \,,
\label{innerproduct}
\end{eqnarray}
we solve for the expansion constant,
\begin{equation}
\label{BI} B^i_m  = \frac{-m}{{\chi_{m1}}^2 {\cal
N}_{{\psi}_{m;1}}} {\bf L}_m \,,
\end{equation}
where ${\cal N}_{{\psi}_{m;1}}$ is a normalization factor given in the
Appendix and the matrix element
\begin{eqnarray}
\label{LM} {\bf L}_m &=& 
\Biggl<
{\cal C}_{m;1}\frac{D{q}^{(0)}}{r} \nonumber \\ 
&\times&~~~ {\biggl[1 + 
\frac{4}{{\chi_{m1}}^2} 
\biggl(
\gamma^i -
\frac{m D{\phi}^{(0)}}{r}\biggr)^2 \biggr]}^{-1}  \nonumber \\ 
&\times&~~~ \psi_{m;1}~\Biggr>.
\end{eqnarray}
A similar projection of Eq.~\ref{APPROXCCR} and some
simplification results in
\begin{equation}
\label{BR} B^r_m  = \frac{2 B^i_m }{{\chi_{m1}}}~\Biggl(\gamma^i -
\frac{m}{ {\cal N}_{{\psi}_{m;1}}}~
 \Bigl< ~{{\psi}_{m;1}}
\frac{D{\phi}^{(0)}}{r}~{{\psi}_{m;1}}\Bigr>\Biggr).
\end{equation}
Equations~\ref{BI}-~\ref{BR} determine the expansion
coefficients of the potential and charge density for a given stream function.
Substitution of Eq.~\ref{APPROXEX} into Eq.~\ref{NDNS_mode}, with $\gamma =
i \gamma^i$ and a Couette shear gives
\begin{eqnarray}
{\beta_{m1}}^4 {\cal C}_{m;1} -i~\frac{{\beta_{m1}}^2}{{\cal P}}
 \biggl(
\gamma^i -  \frac{m D{\phi}^{(0)}}{r}\biggr) 
 \bigl( {\cal V}_{m;1}- 
{\cal U}_{m;1}\bigr) \nonumber \\
 - i~ \frac{m{\cal R}}{r}\biggl( (B^r_m + i
B^i_m ) (2~{\chi_{m1}} D{\Psi}^{(0)}  \nonumber \\
- D{q}^{(0)} )
\psi_{m;1} \biggr) = 0,
\end{eqnarray}
where ${\cal U}_{m;1}$ and ${\cal
V}_{m;1}$ are defined in the Appendix. Projecting the above equation with
$C_{m;1}$ allows us to solve for ${\cal R}$:
\begin{eqnarray}
\label{R}  {\cal R}\,(\alpha, {\cal P}, {{\cal R}e}, m, 
\gamma^i) = \nonumber \\
\frac{ {\beta_{m1}}^4 {\cal N}_{{\cal C}_{m;1}} +
i {\cal P}^{-1} {{\beta^2_{m1}}}\biggl( m {\bf F}_m
- {\gamma^i}{\bf G}_m \biggr)}{ m
(i B^r_m-B^i_m)
\biggl( 2~{\chi_{m1}} {\bf J}_m -  {\bf K}_m \biggr) } \,,
\end{eqnarray}
where
\begin{eqnarray}
{\bf F}_m &=& \Bigl<{\cal C}_{m;1}\frac{D{\phi}^{(0)}}{r}
\Bigl(
{\cal V}_{m;1}- {\cal U}_{m;1}\Bigr)\Bigr> \,, \\
{\bf G}_m &=&  \Bigl<{\cal C}_{m;1}  \Bigl( {\cal
V}_{m;1}- {\cal
U}_{m;1}\Bigr) \Bigr> \,, \\
 {\bf J}_m &=&  \Bigl<{\cal C}_{m;1}\Bigl(\frac{D{\Psi}^{(0)}}{r}\Bigl)
\psi_{m;1}\Bigr> \,,\\
 {\bf K}_m &=& \Bigl< {\cal C}_{m;1}\Bigl(\frac{D{q}^{(0)}}{r}\Bigl)
\psi_{m;1}\Bigr> \,.
\label{matrixdefinitions}
\end{eqnarray}
The normalization factor ${\cal N}_{{\cal C}_{m;1}}$ is given in the
Appendix.

To determine the linear stability boundary for a given radius
ratio $\alpha$, Prandtl number ${\cal P}$ and Reynolds number
${{\cal R}e}$, we solve Eq.~\ref{R} for a sequence of azimuthal mode
numbers $m$, using Mathematica for all integrations.  We first consider some
special cases.

\begin{figure}
\epsfxsize =3.6in
\centerline{\epsffile{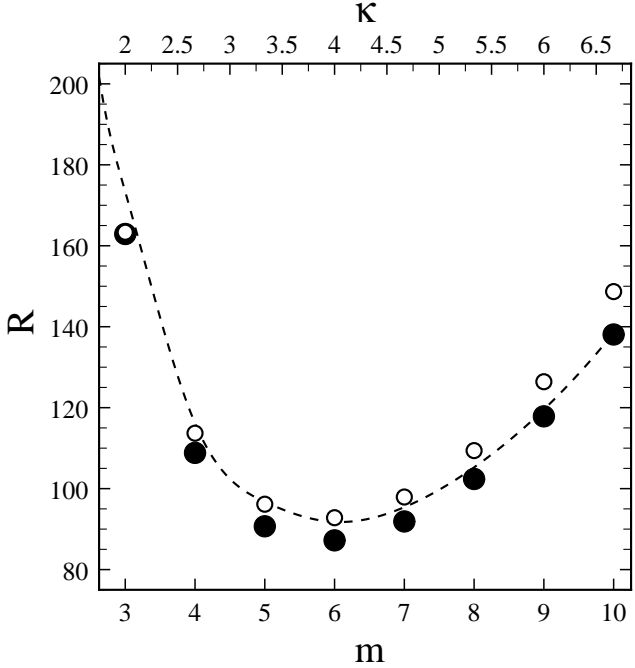}}
\caption{The marginal stability boundary ${\cal R}$ {\it vs.} $m$ for zero
applied shear and $\alpha = 0.5$. Open symbols show the result from the local
approximation (Section~\protect\ref{local}), solid symbols show the  full
nonlocal stability calculation (Section~\protect\ref{nonlocal}), with three
expansion modes. The dashed line indicates the continuous marginal stability
curve for the unbounded rectangular geometry with plate electrodes (see
Ref.~\protect\cite{linear}).  The azimuthal mode number $m$ is equivalent to
a dimensionless wavenumber $\kappa = m/{\bar r}$ (upper scale), where ${\bar
r}$ is the mean radius $(r_i + r_o)/2$.}
\label{mbound_OMzero}
\end{figure}

For zero shear, $\phi^{(0)}= \gamma^i = 0$ and Eq.~\ref{R} reduces to
\begin{equation}
\label{Rzeroshear} {\cal R}\,(\alpha, m) ~=~\frac{{\beta_{m1}}^4
{\chi_{m1}}^2 {\cal N}_{{\cal C}_{m;1}} {\cal N}_{{\psi}_{m;p1}}
}{m^2  {\bf K}_m
\biggl(2~\chi_{m1}{\bf J}_m - {\bf K}_m \biggr)}.
\end{equation}
Note that Eq.~\ref{Rzeroshear} is independent of the Prandtl number
and is always real.  Fig.~\ref{mbound_OMzero} shows the resulting stability
boundary for $\alpha = 0.5$.  Except for being discretized in integer values
of $m$ by the annular geometry, the boundary is not much different than that
of the infinite rectangular   

\noindent case\cite{linear}, which is defined at a
continuum of wavevectors $\kappa$. The discrete curve approaches the
continuum one in the limit $\alpha \rightarrow 1$.  Also shown in
Fig.~\ref{mbound_OMzero} is the more accurate numerical solution presented in
Section~\ref{nonlocal}, below.  The minima of this curve define the
critical parameters $m_c^0 (\alpha) \equiv m_c(\alpha, {\cal R}e = 0)$ and
${\cal R}_c^0 (\alpha) \equiv {\cal R}_c(\alpha, {\cal R}e = 0)$.  Some
values for various $\alpha$ are collected in Table~\ref{criticalparameters}.

For non-zero shear, Eq.~\ref{R} simplifies somewhat in the limit that the
Prandtl number ${\cal P} \rightarrow \infty$. A word of caution is perhaps appropriate here.  As ${\cal P} \rightarrow \infty$, we require  $\Omega \rightarrow \infty $ such that $ \Omega / {\cal P}$ and therefore ${{\cal R}e}$ (see Eq.~\ref{reynolds}) are finite . In this limit, with the proviso
that ${\cal R}$ be real, Eq.~\ref{R} becomes
\begin{equation}
\label{Rinifiniteprandtl}
{\cal R}\,(\alpha, {{\cal R}e}, m,\gamma^i) ~=~\frac{{\beta_{m1}}^4
{\chi_{m1}}^2 {\cal N}_{{\cal C}_{m;1}} {\cal N}_{{\psi}_{m;p1}}}{m^2
{\bf L}_m
\biggl(2~\chi_{m1}{\bf J}_m - {\bf K}_m \biggr)}.
\end{equation}
The only shear dependence in
Eq.~\ref{Rinifiniteprandtl} occurs through the matrix element
${\bf L}_m$, which is defined by Eq.~\ref{LM}.
For zero shear, ${\bf L}_m \equiv {\bf K}_m$, while for
non-zero shear, ${\bf L}_m$ is bounded above by ${\bf K}_m$.  Hence, ${\cal
R}\,(\alpha, {{\cal R}e}, m,\gamma^i)$ is bounded below by
${\cal R}\,(\alpha, {{\cal R}e} = 0 , m,\gamma^i = 0)$, the
zero shear value.
Thus, one expects suppression of the onset of convection for
non-zero shear for every non-axisymmetric mode $m$.  This feature is also
present for finite ${\cal P}$.

\begin{table}
\center
\begin{tabular}{|c|c|c|c|c|c|c|c|}
\multicolumn{1}{|l|}{radius}&
\multicolumn{2}{l|}{local}&
\multicolumn{2}{l|}{nonlocal}&
\multicolumn{3}{l|}{nonlocal}\\
\multicolumn{1}{|l|}{ratio}&
\multicolumn{2}{l|}{($n=1,p=1$)}&
\multicolumn{2}{l|}{($n=1,l=20$)}&
\multicolumn{3}{l|}{($n=3,l=20$)}\\ \hline
{$\alpha$}&{$m^0_c$}&{${\cal R}^0_c$}&{$m^0_c$}&{${\cal
R}^0_c$}&{$m^0_c$}&$m^0_c/{\bar r}$&{${\cal R}^0_c$}\\ \hline
0.33 & 4& 102.58 & 4 & 91.62 &  4 & 4.03& 82.15 \\
0.56 & 7 & 91.25 & 7 & 92.47 &  7 & 3.95&89.64 \\
0.6446 & 9 & 90.04 & 10 & 93.00 & 10 & 4.33&91.12 \\
0.80 & 18 & 88.84 & 19 & 93.59 & 19 & 4.22&93.10 \\
\end{tabular}
\vspace{3mm}
\caption{Critical parameters for zero shear as determined by
the local approximation (Section~\protect\ref{local}) and the full nonlocal
(Section~\protect\ref{nonlocal}) linear stability analysis.
The integer $n$ ($p$ or $l$) is the number of modes (expansion
functions) used in the series representation of the field variables.
The critical wavevector for
electroconvection in a laterally unbounded geometry is $\kappa_c = 4.223$
(see Ref.~\protect\cite{linear}).  The ratio $m^0_c/{\bar r} \rightarrow
\kappa_c$ as $\alpha \rightarrow 1$, where ${\bar r}$ is the mean radius
$(r_i + r_o)/2$.     }
\label{criticalparameters}
\end{table}
\begin{table}
\center
\begin{tabular}{|c|c|c|c|c|c|c|c|}
\multicolumn{1}{|l|}{Prandtl}&
\multicolumn{1}{l|}{Rotation}&
\multicolumn{3}{l|}{local}&
\multicolumn{3}{l|}{nonlocal} \\ 
\multicolumn{1}{|l|}{}&
\multicolumn{1}{l|}{rate}&
\multicolumn{3}{l|}{($n=1, p = 1$)}&
\multicolumn{3}{l|}{($n=1, l = 20$)} \\\hline
{${\cal P}$}&{$\Omega$}&{$m_c$}&{${\cal R}_c$}&{$\gamma_c^i$}&{$m_c$}&{${\cal
R}_c$}&{$\gamma_c^i$} \\ \hline
100  &5.5135& 8 & 319.50 & -14.949 & 8  &  282.61 & -16.383  \\
10   & 0.5513&9 & 99.65 & -1.655 & 9  &  98.52  &  -1.861  \\
1    &0.0551 &9 &  90.16 & -0.166 & 10 &  93.06  &  -0.205   \\
0.1  &0.0055 &9 &  90.05 & -0.017 & 10 &  93.00  &  -0.020  \\
0.01 &0.0006 &9 &  90.04 & -0.002 & 10 &  93.00  &  -0.002  \\
\end{tabular}
\vskip 0.15in
\caption{Critical Parameters for a range of Prandtl numbers ${\cal P}$
at radius ratio $\alpha = 0.6446$ and Reynolds number
${{\cal R}e} = 0.1$.
The integer $n$ ($p$ or $l$) is the number of modes (expansion
functions) used in the series representation of the field variables.}
\label{criticalparametersprandtl}
\vspace{2mm}
\end{table}

\begin{table}
\center
\begin{tabular}{|c|c|c|c|c|c|c|c|}
\multicolumn{1}{|l|}{Prandtl}&
\multicolumn{1}{l|}{Reynolds}&
\multicolumn{3}{l|}{local}&
\multicolumn{3}{l|}{nonlocal} \\ 
\multicolumn{1}{|l|}{}&
\multicolumn{1}{l|}{}&
\multicolumn{3}{l|}{($n=1, p = 1$)}&
\multicolumn{3}{l|}{($n=1, l = 20$)} \\ \hline
{${\cal P}$}&{{\cal R}e}&{$m_c$}&{${\cal R}_c$}&{$\gamma_c^i$}&{$m_c$}&{${\cal
R}_c$}&{$\gamma_c^i$} \\ \hline
100  &0.0181& 9 & 115.20 & -2.995 & 9  &  109.03 & -3.372  \\
10   & 0.1814&9 & 115.29 & -3.002& 9  & 109.02  &  -3.365  \\
1    &1.8137 &9 &  115.27& -3.007 & 9 &  108.99  &  -3.307   \\
0.1  &18.1373 &9 &  115.24 & -3.039 & 9 & 109.14   &  -3.165 \\
0.01 &181.3731&9 &  115.16 & -3.086 & 9 &  109.31  & -3.108  \\
\end{tabular}
\vskip 0.15in
\caption{Critical Parameters for a range of Prandtl numbers ${\cal P}$
at radius ratio $\alpha = 0.6446$ and rotation rate $\Omega = 1.0$.
The integer $n$ ($p$ or $l$) is the number of modes (expansion
functions) used in the series representation of the field variables.}
\label{criticalparametersprandtl2}
\vspace{2mm}
\end{table}

For arbitrary $\alpha$, ${\cal P}$ and ${{\cal R}e}$, one can solve
Eq.~\ref{R} for various $m$ by a 1D search procedure. At each $m$,
$\gamma^i$ is varied to find a ${\cal R}$ that is real and a
minimum.   Example neutral curves are shown in Fig.~\ref{mbound_OM} for
$\alpha=0.8$, ${\cal P}=10$ and several ${{\cal R}e}$. The suppression of
convective onset is evident, as well as a tendency for the critical mode
number $m_c$ to decrease with ${{\cal R}e}$. We find that ${\cal R}({{\cal
R}e})$ is a monotonically increasing function of ${{\cal R}e}$.
The dependence of ${\cal R}$ on ${\cal P}$ is treated in two ways. We will first fix ${{\cal R}e}$, change ${\cal P}$ and consequently let $\Omega$ vary.  In the second protocol we will fix $\Omega$, change ${\cal P}$ and consequently let ${{\cal R}e}$ vary.  For fixed  ${\cal R}e$, the ${\cal P}$ dependence of these results is not
strong, except at large ${\cal P}$.  Because ${\cal R}e$ is proportional to
${\cal P}^{-1}$ in Eq.~\ref{reynolds}, this limit corresponds to large values
of $\Omega$ and a large suppression effect. Some results for various ${\cal
P}$ are summarized in Table~\ref{criticalparametersprandtl}. For fixed $\Omega$, the ${\cal
P}$ dependence is very weak at all ${\cal
P}$ investigated. Results are summarized in Table~\ref{criticalparametersprandtl2}.

\section{Linear Stability: Exact nonlocal solution}
\label{nonlocal}

We now relax the local approximation described in the previous section, and
properly treat the full problem.
The electrostatic Eqs.~\ref{NDLAP_mode}-\ref{NDqdef_mode}, in which the
charge density and electrostatic potential are related nonlocally, are solved
numerically.  As well as being more physically correct, this allows us to
determine the accuracy of the local approximation.

The first step in the present method is to find the appropriate
expansion functions of the field variables. Substitution of
Eqs.~\ref{phi_m}-\ref{q_m} into Eqs.~\ref{NDCC_mode}-\ref{NDqdef_mode}
yields equations which can be solved for $A_n$, $\phi_{m;n}$,
$\Psi_{m;n}$, $\Psi_{3m;n}$, and $Q_{m;n}$. The stream function
${\phi}_{m;n}(r) = C_{m;n}(r)$ as in Eq.~\ref{Cmn}. The potentials
$\Psi_{m;n}$ and $\Psi_{3m;n}$ and the charge density $Q_{m;n}$ are further
expanded as
\begin{eqnarray}
\Psi_{m;n}(r) & = & \sum_l v_{m;nl}
\psi_{m;l}(r) \,,
\label{Psi_mn_OM} \\
\Psi_{3m;n}(r) & = & \sum_l v_{m;nl}
\psi_{3m;l}(r) \,,
\label{Psi_3mn_OM} \\
Q_{m;n}(r) & = & \sum_l v_{m;nl} q_{m;l}(r) \,,
\label{q_mn_OM}
\end{eqnarray}
where $\psi_{m;l}$ is given by Eq.~\ref{psimn} and $v_{m;nl}$ are complex
coefficients. The functions $\psi_{3m;l}$ and $q_{m;l}$
are computed as follows. After substituting Eqs.~\ref{Psi_3m}
and~\ref{Psi_3mn_OM} into Eq.~\ref{NDLAP_mode}, the resulting equation
\begin{eqnarray}
\biggl( D_*D - \frac{m^2}{r^2} + \frac{\partial^2}{\partial z^2} \biggl)
  {\psi}_{3m;l} & = & 0 \,, \label{Helmholtz}
\end{eqnarray}
is solved numerically on a finite 2D grid by an over-relaxation
algorithm\cite{recipes}
for the functions $\psi_{3m;l}(r,z)$, subject to the boundary conditions as
in Eqs.~\ref{directbc3mn}-\ref{directbc4mn}
with $\psi_{3m;l}(r,0) =\psi_{m;l}(r)$ for $r_i \leq r \leq r_o$ and $0$
otherwise. Then Eqs.~\ref{NDqdef_mode},~\ref{Psi_3m},~\ref{q_m}, and
{}~\ref{Psi_3mn_OM}-\ref{Helmholtz} give

\begin{figure}
\epsfxsize =3.7in
\centerline{\epsffile{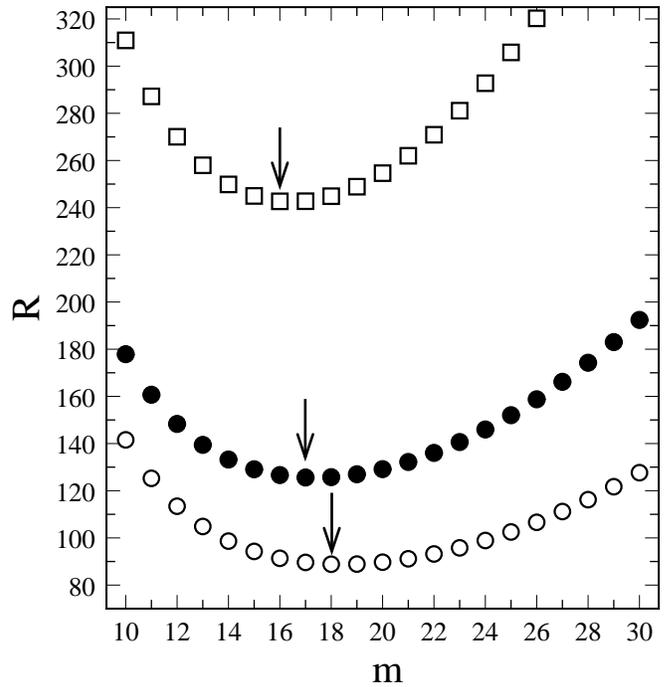}}
\vskip 0.15in
\caption{Marginal stability boundaries under shear, showing ${\cal
R}$ as a function of $m$ for $\alpha=0.80$ and ${\cal P} = 10$, using the
local approximation (Section~\protect\ref{local}).  Shown are the boundaries
for Reynolds numbers ${{\cal R}e}=$0.0 (circles), 0.25 (filled circles) and
0.75 (squares).  Arrows indicate the critical mode.}
\label{mbound_OM}
\end{figure}
\vspace{4mm}

\begin{equation}
q_{m;l}(r) = -2 \partial_z \psi_{3m;l}(r,z) |_{z=0^+} \,, \label{qmn}
\end{equation}
where the differentiation is performed numerically.
Fig.~\ref{qcomparison} shows a plot of the charge density $q_{6;1}$ at
$\alpha = 0.5$, and compares the numerical result to the approximate one
given by Eqs.~\ref{QWF} and~\ref{f} from the previous section. The
approximate solution does not contain the divergences which occur near the
film's edges, due to the sharp changes in the derivative of the potential.
These
are a feature of our 2D model, which treats the film and electrodes as having
zero thickness.

We next substitute Equations~\ref{Psi_mn_OM} and
\ref{q_mn_OM} into Eq.~\ref{NDCC_mode} and use
\begin{equation}
C_{m;n} \bigg(\frac{Dq^{(0)}}{r}\bigg) = \sum_l
\Biggl< C_{m;n} \bigg(\frac{Dq^{(0)}}{r}\bigg) \psi_{m;l} \Biggr>
 \psi_{m;l} \,.
\label{mDphi_expan}
\end{equation}
The resulting equation is projected against $\psi_{m;k}$ to obtain a matrix
expression that can be solved numerically for the complex coefficients
$v_{m;nl}$.

Finally, substituting the various expansions in $\psi_{m;l}$ into
Eq.~\ref{NDNS_mode} and taking the inner product
with $C_{m;p}$, yields a set of linear homogeneous equations for the
constants $A_n$.  We write this set
as the matrix equation $\sum_n A_n {\bf T}_{pn}=0$.
For a nontrivial solution, the compatibility condition is
\begin{equation}
 {\rm Real}(det~[{\bf T}])~=~{\rm Imag}(det~[{\bf T}])~=~0 \,,
\label{compat_OM}
\end{equation}
with the elements of the matrix ${\bf T}$ given by
\begin{eqnarray}
{\bf T}_{pn} = \Biggl< C_{m;p} \biggl(D_{*}D -  \frac{m^2}{r^2}
\biggr)^2 C_{m;n}
\Biggr>
 \nonumber \\
  - \frac{i}{{\cal P}} \Biggl< C_{m;p} \biggl(\gamma^i -
m\bigg(\frac{D\phi^{(0)}}{r}\bigg)
\biggr) \biggl(D_{*}D - \frac{m^2}{r^2} \biggr) C_{m;n} \Biggr> \nonumber \\
  - i m{\cal R} {\bf F}_{m;pn}
 \,.  \label{Tpn_OM}
\end{eqnarray}
The first two inner products of Eq.~\ref{Tpn_OM} can be simplified using
Eqs.~\ref{eigeneq1} and \ref{eigeneq2} and the matrix elements
${\bf F}_{m;pn}$ are
\begin{equation}
{\bf F}_{m;pn}  =  \sum_l v_{m;nl}
 \biggl< \frac{C_{m;p}}{r} \biggl( (D\Psi^{(0)}) q_{m;l}
 - (Dq^{(0)})\psi_{m;l} \biggr) \biggr> \,.
\end{equation}
The real values of ${\cal R}$ and $\gamma^i$ which satisfy Eqs.~\ref{compat_OM}
and~\ref{Tpn_OM} at each $m$ define the neutral stability boundary ${\cal
R}={\cal R}\,(\alpha, {\cal P}, {{\cal R}e}, m, \gamma^i)$.  The critical
parameters $m_c$, ${\cal R}_c$ and $\gamma_c^i$ are obtained when ${\cal R}$
is minimized.

The numerical over-relaxation calculation used to solve Eq.~\ref{Helmholtz}
involved a grid spacing such that there was a minimum of 160 points across the
width of the film. For the purposes of integration, the discrete values of
$q_{m;l}$ found numerically from Eq.~\ref{qmn} were Chebyshev interpolated.
The series in Eqs.~\ref{Psi_mn_OM}-\ref{q_mn_OM} and \ref{mDphi_expan} were
calculated up to $l=20$. Three modes ($n=p=3$)
were employed in the compatibility conditions Eq.~\ref{compat_OM} when the
shear
was zero. This was reduced to one mode ($n=p=1$) when the shear was non-zero.
All $r$-integrations were performed by the Romberg method.\cite{recipes}
We solved ${\rm Real}(det~[{\bf T}])=0$ (Eq.~\ref{compat_OM}) for ${\cal R}$
for a given trial $\gamma^i$ and associated coefficients $v_{m;nl}$ of the
field variables. The ${\cal R}$ and trial $v_{m;nl}$ were employed in the
search for the $\gamma^i$ which satisfied ${\rm Imag}(det~[{\bf T}])=0$.
The new value of $\gamma^i$ was then used to find new coefficients
$v_{m;nl}$. The iterative cycle was continued until the parameters and
field variables were converged.

The marginal stability boundary for $\alpha = 0.5$ and zero shear is shown
in Fig.~\ref{mbound_OMzero}, using both the local approximations and the full
nonlocal solution. Neither solution is qualitatively much different from the
infinite rectangular case.\cite{linear} For the nonlocal solution, we find
that as $\alpha \rightarrow 1$, $m_c^0$ increases such that $m_c^0/{{\bar
r}}$ approaches the correct limiting value, $\kappa_c = 4.223$. Here, ${\bar
r}=(r_i+r_o)/2$ is the midline radius of the annulus and $\kappa_c$ is the
critical wavenumber for an infinite rectangular film in the ``plate''
electrode geometry.\cite{linear} Table~\ref{criticalparameters} shows a
sample set of critical parameters as determined by both solution schemes.
There is generally good agreement between the two methods, except at small
$\alpha$, where more expansion modes are needed.

\begin{figure}
\epsfxsize =3.7in
\centerline{\epsffile{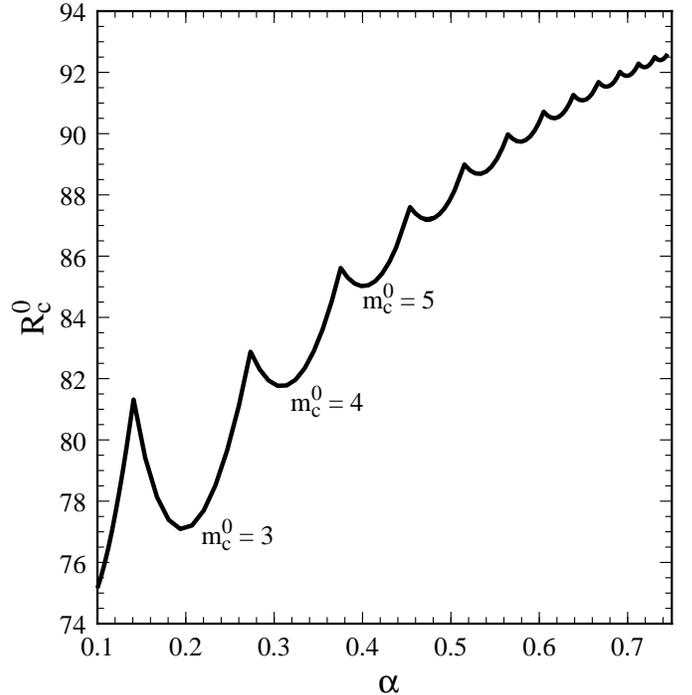}}
\vskip 0.2in
\caption{The critical control parameter for zero shear ${\cal R}_c^0$ versus
radius ratio $\alpha$ using the nonlocal stability analysis with three
expansion modes. Special radius ratios, at which two adjacent modes $m$
and $m+1$ are simultaneously marginally unstable, occur at the
cusps of the curve. Between cusps, the critical mode number $m_c^0$ remains
constant.}
\label{scallops_OMzero}
\end{figure}

Fig.~\ref{scallops_OMzero} shows the zero shear critical control parameter
${\cal R}_c^0$ as a function of the radius ratio $\alpha$. There are discrete
values of $\alpha$ where ${\cal
R}_c^0\,(\alpha,m_c)={\cal R}_c^0\,(\alpha,m_c + 1)$, so that

\noindent two adjacent
modes become unstable simultaneously at a co-dimension two point.
These points occur at the cusps in
Fig.~\ref{scallops_OMzero}, while between the cusps a single
value of $m$ is critical. It is interesting to note that the trend in ${\cal
R}^0_c$ is increasing overall, as function of $\alpha$. This is opposite to
what is found for radially driven Rayleigh-B{\'e}nard convection in a
rotating annulus.\cite{alonso95}  We attribute this difference as a manifestation of the differences between the nature of the charge density `inversion' and thereby electrical forcing with the bouyancy inversion of RBC.  As $\alpha \rightarrow 1$, the
co-dimension two points become closely spaced and the value of ${\cal R}_c^0$
approaches a limiting value. \cite{extrap_foot}

For the case of non-zero shear, the nonlocal analysis produces neutral curves
which resemble those shown in Fig.~\ref{mbound_OM}.  For fixed $\alpha$ and
${\cal P}$, the critical control parameter ${\cal R}_c$ increases with
${{\cal R}e}$, which indicates a suppression of convection as is
observed in experiments.\cite{annular98}  This prediction is compared to
experimental data in Section~\ref{expmethod} below.  The ${\cal P}$
dependence of ${\cal R}_c$ for ${{\cal R}e \neq 0}$ is rather weak at small
${\cal P}$. This is outlined in Tables~\ref{criticalparametersprandtl} and \ref{criticalparametersprandtl2}. Note that for fixed ${{\cal R}e}$, large ${\cal P}$ implies large $\Omega$ and thus greater stabilization.  For fixed $\Omega$, however, the stability propoerties are insensitive to ${\cal P}$ and therefore to ${{\cal R}e}$.

\begin{figure}
\epsfxsize =3.5in
\centerline{\epsffile{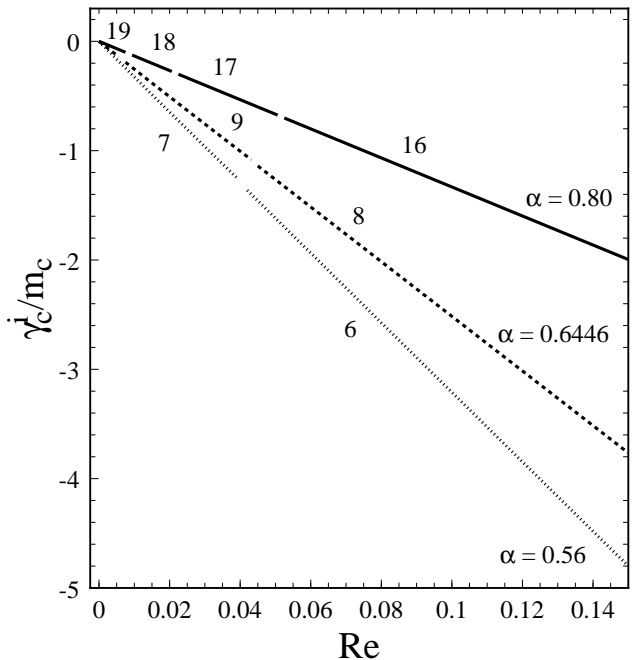}}
\caption{Ratio of imaginary part of the growth rate to critical
mode number, $\gamma_c^i/m_c$ versus Reynolds number ${{\cal R}e}$ for
$\alpha=$0.56, 0.6446, and 0.80 and ${\cal P} = 123$ using
the nonlocal linear stability calculation with one expansion mode.
The breaks in the curves for each $\alpha$ show the intervals over which the
critical azimuthal mode number $m_c$ has the value indicated.}
\label{gamma_i_plot}
\end{figure}

The critical mode number $m_c$ decreases with increasing shear.  This effect
is shown in Fig.~\ref{gamma_i_plot}, which displays 

\noindent the angular traveling rate
of the critical mode, $\gamma^i_c/m_c$, as a function of ${{\cal R}e}$ for
several $\alpha$.  For $\Omega > 0$,  $\gamma^i_c < 0$, which indicates that
the critical mode travels around the annulus in the same sense as the inner
electrode. $\gamma^i_c/m_c$ is a very nearly linearly decreasing function of
${{\cal R}e}$, with very small discontinuities at points where the critical
azimuthal mode number $m_c$ changes, as shown in Fig.~\ref{gamma_i_plot}.
Each of these discontinuities, which are too small to resolve on the scale of
Fig.~\ref{gamma_i_plot}, is a co-dimension two point, where two adjacent $m$
modes with very slightly different traveling rates are simultaneously
unstable at onset.  These special points are also slightly ${\cal P}$
dependent, as well as being $\alpha$ dependent in a manner similar to the
zero shear case discussed above.

\begin{figure}
\epsfxsize =3.3in
\centerline{\epsffile{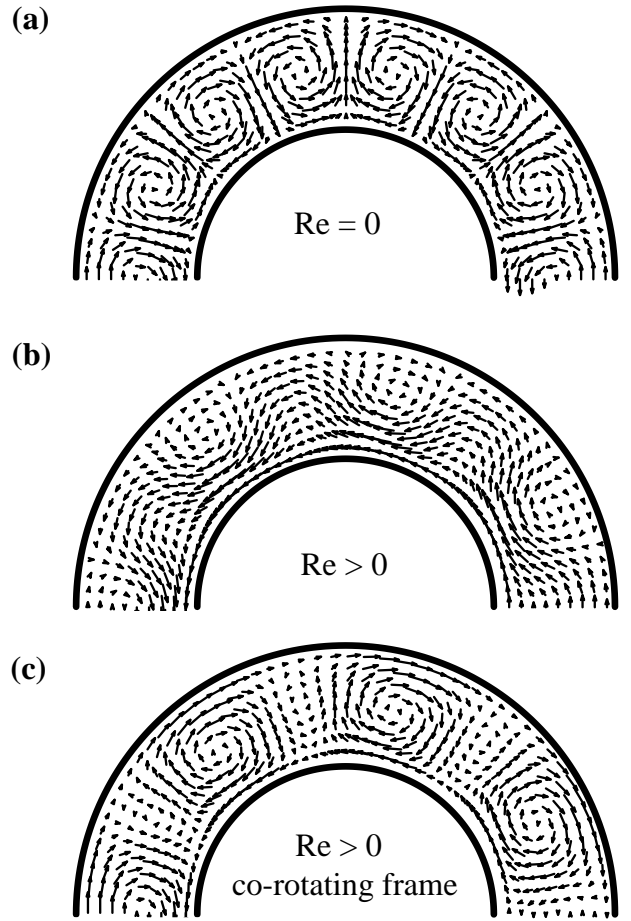}}
\vskip 0.15in
\caption{The velocity vector field of arbitrary amplitude for
$\alpha=0.56$ with (a)~${\cal R}e = 0$, (b)~${\cal R}e > 0$ as viewed in the
laboratory frame, and (c)~${\cal R}e> 0$ as seen in the frame co-rotating at
$\gamma^i_c/m_c$ in which the pattern is stationary.}
\label{vectorplot}
\end{figure}

Although linear analysis cannot provide the magnitude of the fields above
onset, it is nevertheless interesting to examine the spatial structure of the
linearly unstable modes. Figure~\ref{vectorplot} displays the velocity
vector field of a critical mode for $\alpha=0.56$, plotted with an arbitrary
amplitude.   In Fig.~\ref{vectorplot}(a), we show the stationary vortex
pattern at ${{\cal R}e}=0$.  Here $m_c^0 = 7$, so 7 symmetric vortex pairs are
arranged around the annulus.  This pattern is purely non-axisymmetric or ``columnar''.  It is an exact solution to the governing equations and has been observed experimentally.  As is discussed in Section~\ref{discussion} below, we would expect exactly the same pattern under the same conditions for any rate of rigid rotation.  In Fig.~\ref{vectorplot}(b), we show the
typical flow pattern for 

\noindent a large ${{\cal R}e}>0$, as viewed in the laboratory
frame.  The periodicity of the pattern is reduced ({\it i.e.} $m_c <
m_c^0$), and the traveling pattern appears as a meandering wave in the
laboratory frame.  Fig.~\ref{vectorplot}(c) shows the same ${{\cal R}e}$ as
in (b), but as seen in the frame in which the pattern is stationary.  This
frame rotates in the same sense as the  inner electrode but with an angular speed
of $ \gamma^i_c/m_c$, which is less than $\Omega$. In this frame, each
vortex pair consists of a larger and a smaller member. The same qualitative
features are seen in sheared convection patterns just above onset in the
present experiments and those of Ref.~\cite{annular98}.  

\section{Experiment}
\label{expmethod}

In this section, we discuss our experimental
investigation of the linear stability of 2D fluids to electroconvection
and Couette shear.  The apparatus we describe is only slightly modified
from that previously reported in Ref~\cite{annular98}.  Our objective
here is to experimentally test one of the direct predictions of the
linear theory, namely the suppression of onset by shear.

A smectic~A liquid crystal (octylcyanobiphenyl, or 8CB) was employed to
realize a 2D fluid.  In the smectic~A phase, long organic molecules
arrange themselves into layers with their long axes normal to the layer.
Within the layer planes, smectic~A behaves as a 2D isotropic fluid, and
can be drawn into robust submicron films which are integer numbers of
layers thick and uniform over their whole area.\cite{Morris,LC} Other
smectic phases form 2D anisotropic fluid
films.\cite{LC,smcexp,smctheory} Unlike soap films,
smectics can flow in the film plane without thickness change and are not
susceptible to evaporation.  Smectics generally have low electrical
conductivity due to residual ionic impurities.\cite{LC,8cb}  This
combination of properties makes smectic films uniquely suited for
studying electroconvection.  While our experiments used liquid crystal
films,
the electroconvective instability itself is generic and not
specific to anisotropic fluids; a similar instability is found in thin
drops or puddles of ordinary isotropic fluids.\cite{faetti,avsec,malkus}
The anisotropy introduced by the smectic layering simply serves to
constrain the film to have a constant, very small, thickness.  It behaves
isotropically for flows in the plane of interest.

Our experimental apparatus consisted of two concentrically placed
stainless steel electrodes which supported a suspended film. The inner
electrode was a disk of radius $3.61 \pm 0.01$~mm and the outer electrode
was a plate with a circular hole of radius $6.40 \pm 0.01$~mm, giving an
annular film with radius ratio $\alpha = 0.564 \pm 0.002$.  Each
molecular layer of smectic~A~8CB is $3.16$~nm thick.  The inner electrode could
be
rotated about its axis by a high precision stepper motor at angular
frequencies up to $4$~rad/s, while the outer was fixed.  All experiments
were performed at $23 \pm 1^\circ$C.  The air surrounding the film was
pumped down to an ambient pressure of 0.5-5 torr.  The low pressure ensures
that air drag on the film will be insignificant.\cite{airdrag}  The
system was electrically shielded by a Faraday cage and the outer
electrode and shield were maintained at ground
potential while the voltage $V$ of the inner electrode was varied above
ground. The resulting current through the film was measured using a
picoammeter.  Except for a few qualitative runs to observe the vortex
patterns, we did not visualize the flow, as this required suspending
particles in the film, which was not necessary for our present purposes.
The film behaved as an ohmic conductor in the conduction regime below the
onset of convection. The critical voltage $V_c$, at which the onset of
convection  occurred, was located from the sharp increase in current. In
all runs, the shear rate was held constant while the voltage was slowly
increased and decreased through onset. For some rates of shear, the onset
of convection was hysteretic.\cite{annular98} In these cases, we took
$V_c$ to be the transition from conduction to convection as $V$ is increased.

In order to interpret the experimentally measured quantities, we need to
accurately express them in nondimensional form.  According to
Eq.~\ref{Rayleigh}, the  critical control parameter ${\cal R}_c$ is
related to the critical voltage $V_c$ by
\begin{equation}
{\cal R}_c =
\frac{\epsilon_0^2 {V_c}^2}{\sigma_3 \eta_3 s^2} \,. \label{Rayleigh_c}
\end{equation}
The size of the experimental uncertainties on $\sigma_3$, $s$, and
especially, $\eta_3$ prevented us from directly using
Eq.~\ref{Rayleigh_c} to scale the voltages.  Rather, we used the zero
shear measurements of $V_c^0$ to scale the data for non-zero shear.  A
relative measure of the suppression of the onset of convection is
\begin{equation}
\tilde{\epsilon}~({{\cal R}e}) = \biggl[\frac{{\cal R}_c({{\cal
R}e})}{{\cal R}_c^0}\biggr]-1 = \biggl[\frac{V_c({{\cal
R}e})}{V_c^0}\biggr]^2-1 \,.
\label{epsplus}
\end{equation}
The first part of Eq.~\ref{epsplus} was applied to the theoretical
results, while the second was applied to the experimental data. The latter
scaling does not require a knowledge of the material parameters. In order
to obtain the remaining dimensionless quantities, ${\cal R}e$ and ${\cal
P}$, estimates of $\sigma_3$, $s$, and $\eta_3$ are required.
However, we will show below that our final results are independent of $s$
and $\sigma_3$ and quite insensitive to the value of $\eta_3$.

It is straightforward to measure a quantity proportional to the product
$\sigma_3 s$ from the data. For $V<V_c$, the film behaves ohmically and
hence the slope of the current-voltage data is the film conductance $c$,
given by
\begin{equation}
c = \frac{2\pi \sigma_3 s}{\ln( 1/\alpha )} \,.
\label{conductance}
\end{equation}
For a given current-voltage sweep, $c$ can be measured to within a few
percent.  This precision is limited by the tendency of the conductance of the
material to drift over the course of the sweep.\cite{drift}  Using many
measured values of $c$, we found an average value of $\sigma_3$, which
was consistent with all the data.  The film thickness $s$ can be
independently deduced by examining the color of the film under reflected
white light.\cite{MDMgle}  By measuring $c$ and $s$ for a number of
films with various $\alpha$, we found the drift-averaged conductivity to
be $\sigma_3 =1.3 \pm 0.2 \times 10^{-7} (\Omega m)^{-1}$.  We used this
value to infer a consistent value of $\eta_3$, via the procedure
described below.

Eqs.~\ref{Rayleigh_c} and~\ref{conductance} can be combined to give
\begin{equation}
\Biggl[\frac{2\pi \epsilon_0 }{\ln( 1/\alpha ) \sqrt{{\cal
R}_{c}^0}}\Biggr] V_c^0  = \sqrt{\frac{\eta_3}{\sigma_3}}~c \,.
\label{scaledVc}
\end{equation}
To find $\eta_3$, we plotted the left hand side of Eq.~\ref{scaledVc} against
measurements of $c$, using measured values of $V_c^0$ and the theoretical
value of ${{\cal R}_{c}^0}$.  We used data from a large number of films with
$s < 60$ smectic layers for two values of $\alpha$, 0.56 and 0.64.
A one parameter fit to find the slope of this plot established
$\sqrt{{\eta_3}/{\sigma_3}}$.  Using the average value of
${\sigma_3}$ given above, we found  $\eta_3 = 0.16 \pm 0.06 $~kg/ms.   This
method has the advantage that it does not require an independent
measurement of $s$ for every film.

\begin{figure}
\epsfxsize =3.4in
\centerline{\epsffile{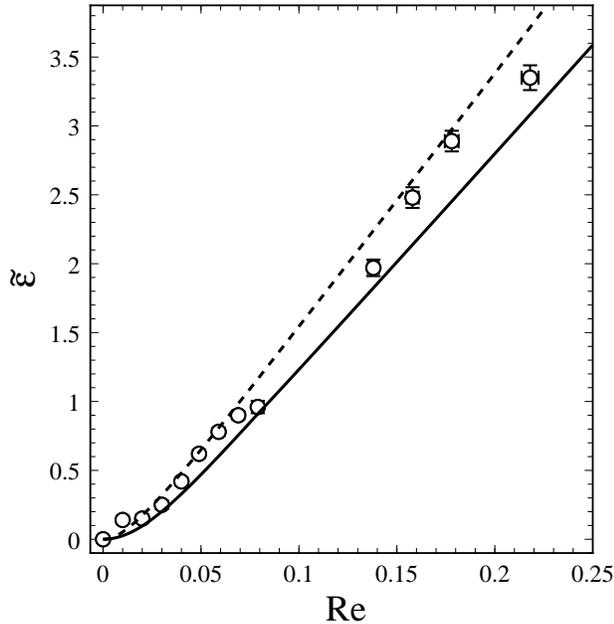}}
\caption{Comparison of the experimental and theoretical relative suppression
$\tilde{\epsilon}$ versus
Reynolds number ${{\cal R}e}$ for
$\alpha=0.56$.  The dashed (solid) line is the result of the local
(nonlocal) method. }
\label{suppression_plot}
\end{figure}

We then employed the above value of $\eta_3$ to calculate ${\cal R}e$  which
is expressed in terms of dimensional parameters by
\begin{equation}
{{\cal R}e} = \rho_3  \omega r_i (r_o -
r_i) / \eta_3~,  \label{reynoldsdimensional}
\end{equation}
where $\omega$ is the measured angular frequency of the inner electrode
in rad/s.  The 3D density $\rho_3$ of the liquid crystal
is\cite{8CBdensity} $1.0 \times 10^{3}$ kg/m$^3$.  Similarly, the Prandtl
number ${\cal P}$ is given in terms of measured quantities by
\begin{equation}
\label{pdimensional} {\cal P} = \frac{2\pi \epsilon_0 \eta_3}{\rho_3
(r_o - r_i) \ln{( 1/\alpha )}}~~\frac{1}{c}\,.
\end{equation}
 From the definition, Eq.~\ref{Rayleigh}, ${\cal P}$ depends on
$s$. In Eq.~\ref{pdimensional}, this dependence is contained in $c$,
which can be accurately measured for each individual current-voltage
sweep.  Thus, for films of various constant thicknesses, ${\cal
P}$ experiences a small drift from sweep to sweep, due to the drift in $c$.
For the data we report here,  ${\cal P}$ was always in the range
$52 < {\cal P} < 75$.  If $c$ is treated as an independent, experimentally determined variable, then Eq.~\ref{pdimensional} is independent of $s$ and $\sigma_3$.

We can now compare our experimental results and the
predictions of the linear theory for non-zero shear.
Fig.~\ref{suppression_plot} shows the theoretical curves from the local
and nonlocal calculations and the
experimental data for the relative suppression $\tilde{\epsilon}$, as a
function of ${\cal R}e$.  The theory curves are plotted for a fixed
${\cal P}=63.5$.  There are no adjustable parameters.  We find good
agreement between the experimental data and the linear theory for ${\cal
R}e < 0.22$.  The local approximation slightly overestimates the
suppression, while the exact nonlocal result slightly underestimates it.
Recall that the local approximation neglects the divergence in the perturbed
charge density at the edges of the film, but uses the full base state
solution, which has similar divergences in our 2D model.  The nonlocal
calculation keeps the divergences at both orders.  The fact that these two
treatments bracket the data suggests that the divergences are rounded off by
the finite thickness of the film and
electrodes in the actual experimental geometry.

The good agreement shown in Fig.~\ref{suppression_plot} is essentially
independent of the value of $\eta_3$.
Since the $\eta_3$ dependence in the ${\cal
R}e$ scaling of both the theory ({\it via} ${\cal P}$ in
Eq.~\ref{reynolds}) and the experiment (according to
Eq.~\ref{reynoldsdimensional}) are proportional to $1/\eta_3$, any
change in $\eta_3$ multipies both by the same factor.  This simply
results in a rescaling of the ${\cal R}e$ axis in
Fig.~\ref{suppression_plot}, with no change in the quality of the
agreement.

\section{Discussion}
\label{discussion}

It is interesting to compare our results for 2D electroconvection under shear with those for other, more familiar instabilities under shear or rotation. 

In a theoretical study, Agrait and Castellanos\cite{agrait88} considered electroconvection due to a radial field in a 3D concentric cylinder geometry with charge injection on either cylinder. Both cylinders were permitted to rotate to produce a general Couette shear. Their result was that shearing enhanced the instability, leading to a 3D flow that resembles TVF.  This can easily be understood as a consequence of the fact that in 3D, the base state Couette shear itself is unstable.  In our 2D system, the shear flow is not only stable by itself\cite{drazin,wmkg95}, but, as we have shown, has a stabilizing effect on the electroconvection.  This is one of several examples of how the 2D nature of our flow leads to an unusual stability, which is often difficult or impossible to realize experimentally in other systems.  

The added stability in our system is a consequence of the shear and not of rotation.  Under rigid rotation, where the inner and outer electrodes are co-rotating, one can transform to rotating co-ordinates in which the electrodes are stationary.  This transformation introduces a coriolis term $-2\Omega \hat{\bf z} \times {\vec{\bf v}} = -2\Omega \nabla \psi $ in Eqn.~\ref{NS} which may be absorbed into the pressure gradient term $\nabla P$ and eliminated.\cite{alonso95} Thus, in a purely 2D system, rigid rotation and the non-rotating, unsheared case have identical stability.  It also follows, since the transformation is general and the unsheared bifurcation is stationary, that the resulting nonlinear vortex pattern above onset must be stationary in the co-rotating frame.

This lack of dependence under rigid rotation may be contrasted with a large class of 3D and quasi-2D rotating Rayleigh-B{\'e}nard systems\cite{dishpan,goldstein9394,rotRBC,knobloch}, where rotation produces added stability but the absence of strictly 2D flow results in time-dependence (precession) of the convection pattern in the co-rotating frame. Chandrasekhar\cite{Chandrasekhar} treated the classic problem of the linear stability of RBC in a laterally unbounded layer rotating about its normal.  The case of a laterally bounded cylindrical layer has received much recent interest theoretically\cite{goldstein9394}, and has also been the subject of a precise experimental study\cite{rotRBC}.

A concise summary of some of the work on rotating RBC can be found in Ref.~\cite{knobloch}.  Due to the similar symmetries in these scenarios, some results are common to most rotating RBC systems.\cite{knobloch}  Since, these systems are three-dimensional (3D), the solutions they support can generally be classified into axisymmetric, non-axisymmetric (``columnar'') or mixed (combinations of axisymmetric and non-axisymmetric) solutions.  The principal result is that for 3D mixed solutions, the onset bifurcation is no longer steady as it is for non-rotating RBC.  This leads to a flow pattern that precesses in the co-rotating frame.  The onset of these solutions is however suppressed, raising the critical Rayleigh number above its non-rotating value.\cite{goldstein9394,rotRBC,knobloch}  Purely non-axisymmetric or columnar solutions are strictly two-dimensional (2D). When they do occur, they do not precess but their onset occurs at the same critical Rayleigh number as in the absence of rotation.\cite{alonso95,knobloch} There have also been theoretical studies of the interesting but nonetheless experimentally unrealizable situation of 2D RBC in a rotating annular geometry with purely {\em radial} gravity and heating\cite{alonso95}. These studies found similar columnar solutions.  In fact, purely columnar solutions (``Taylor columns''\cite{taylor_columns}) have yet to be observed in any rotating RBC experiment since the boundary conditions at the top and botton of the cylinder must be stress free\cite{alonso95,knobloch}, a requirement that cannot be attained in terrestrial RBC experiments.   In contrast, two-dimensionality, stress free end boundary conditions and radial driving forces all arise naturally in the electroconvection of an annular suspended film that we have described.

We next consider sheared, but non-rotating systems.  RBC has been studied experimentally and theoretically with an open through-flow.\cite{ch93,rbc_flow}  The through-flow is generally a weak Poiseuille flow with a very small Reynolds number. Its effects on RBC are well understood.  In brief, the onset of convection is again suppressed, but the first instability is {\it convective} ({\it i.e.} it grows only downstream of a localized perturbation), rather than absolute. The resulting convection pattern drifts in the direction of the through flow.  It is interesting that the clear distinction between convective and absolute instability is blurred in systems like ours, in which the ``through'' flow loops back on itself.  As we have analyzed it, the onset of convection in our annular electroconvection is absolute.

There have been some theoretical studies of 3D RBC in the presence of a {\it plane} Couette shear flow\cite{fk88}, another situation which is experimentally unrealizable. Here, linear analysis reveals stability differences between transverse roll disturbances, those with axes are perpendicular to the shear, and longitudinal roll disturbances which have axes parallel to the shear flow.  Longitudinal-roll disturbances have identical stability properties to RBC in the absence of shear, and are always more unstable than the transverse-roll disturbances.  Transverse-roll disturbances, conversely, exhibit suppression, or added stability due to the shear.  The onset Rayleigh number of transverse rolls is a monotonically increasing function of the shear Reynolds number, similar to what we find for 2D electroconvection.   Furthermore, the critical wavenumber of the most unstable transverse disturbance was found to be a monotonically decreasing function of the shear Reynolds number, also as we observe.  Transverse rolls (vortices, in fact) appear at onset in our 2D system, but would always be pre-empted by longitudinal rolls in a 3D RBC experiment, if it could be realized.

\section{Summary and Conclusion}
\label{conclusion}

In this paper, we have analyzed the linear stability of a weakly
conducting, two-dimensional annular fluid to radially-driven
electroconvection. We considered the effects of an imposed Couette shear
produced by the
rotation of the inner edge of the annulus. We calculated the marginal
stability boundary for the Rayleigh number ${\cal R}(\alpha, {\cal P},{{\cal
R}e},m)$ as a function of the azimuthal mode number $m$, for general radius
ratio $\alpha$, Prandtl number ${\cal P}$ and Reynolds number ${{\cal R}e}$.
We also found the azimuthal traveling rate $\gamma^i/m$ of the marginally
unstable modes, where $\gamma^i$ is the imaginary part of the linear growth
rate. We found a set of discrete values of $\alpha$, ${\cal P}$ and ${{\cal
R}e}$ for which there are two adjacent azimuthal modes $m$ which are
simultaneously unstable at onset. Two solutions schemes were employed: a local
approximation which neglected some aspects of the coupling of fields and
charges, and a more exact, fully nonlocal method. These methods agreed,
except in some quantitative details.

When there is no applied shear, so that ${{\cal R}e}=0$, we found
$\gamma^i=0$ and the most unstable linear mode did not travel. As $\alpha
\rightarrow 1$, the marginal stability boundary for this case approached that
for a laterally unbounded rectangular film.\cite{linear} For non-zero shear,
${{\cal R}e} \neq 0$, the most unstable linear mode traveled in the
direction of the rotation of the inner electrode and the onset of convection
was suppressed, relative to the zero shear case.

  The shear also reduced the
critical azimutal mode number $m_c$.  Rigid rotation has no effect on the critical azimuthal mode number $m_c$, again contrasting with the class of 3D rotating Rayleigh-B{\'e}nard systems\cite{rotRBC} where increasing rotation rate favored  {\em higher} critical azimutal mode number $m_c$.

We performed experiments on thin, suspended annular films of smectic~A
liquid crystals, which are well described by our theoretical model.  We
measured the current through the film as a function of the driving voltage for
a fixed rotation rate of the inner electrode.  This data was used to
determine the onset of convection and two of the material parameters needed
to scale the data.  We compared the relative suppression of the onset from
the experiments to that predicted by the linear theory and found good
agreement.  The spatial structure of the traveling modes from the linear
theory resembled the weakly nonlinear vortex patterns observed
experimentally, when viewed in a co-rotating reference frame.\cite{annular98}

Due to the rich nonlinear aspects of this system, there are several interesting future directions. The system is
sufficiently tractable that the weakly nonlinear regime should be accessible
to theoretical investigation.  It may also be interesting to simulate the
nonlinear regime numerically.  On the experimental side, we expect to be able
to study the secondary bifurcations quantitatively, and make good connection
to the nonlinear theory. It would be useful to determine the material
parameters, particularly the viscosity, by separate experiments. It should be
possible to study regimes of higher ${{\cal R}e}$.  Finally, it would be interesting to examine other flows, such
as a time-dependent oscillatory shear, superposed on electroconvection.

\acknowledgements

We would like to thank John R. de Bruyn and Wayne A. Tokaruk for numerous
discussions.  This research was supported by the Natural Sciences and
Engineering Research Council of Canada.

\appendix
\section{}
\subsection{Expansion functions for the stream function}
\label{appendix_stream}
Since the stream function is constrained by rigid boundary
conditions at $r =
r_i$ and $r = r_o$ and obeys Eq.~\ref{NDNS_mode}, it can be
expanded in the eigenfunctions of the square of the Laplacian
operator.\cite{Chandrasekhar}  Hence, the eigenfunctions sought are defined
by the eigenvalue relation
\begin{equation}
\Biggl(D_{*}D - \frac{m^2}{r^2} \Biggr)^2 {\cal C}_{m;n} =
\beta_{mn}^4 {\cal
C}_{m;n} \,.
\label{eigeneq1}
\end{equation}
 The boundary conditions are ${\cal C}_{m;n}(r_i) = {\cal
C}_{m;n}(r_o) = 0$
and $D{\cal C}_{m;n}(r_i) = D{\cal C}_{m;n}(r_o)= 0$.  The desired
solutions of
Eq.~\ref{eigeneq1} are
 \begin{eqnarray}
{\cal C}_{m;n}(r) = J_m(\beta_{mn} r) +B_{mn}
Y_m(\beta_{mn} r) \nonumber \\
+C_{mn} I_m(\beta_{mn} r) +D_{mn} K_m(\beta_{mn} r) \,,
\label{properstream}
\end{eqnarray}
where $J_m$ and $Y_m$ are the familiar Bessel and Weber functions
of order $m$,
$I_m$ and $K_m$ are the modified Bessel functions of order $m$.
The parameters
${\beta_{mn}}$ are successive solutions of the secular equation $det~{\bf
M}(\beta) = 0$, where
\begin{eqnarray}
{\bf M}(\beta) = \nonumber \\
\left[\begin{array}{cccc}
J_m(\beta r_i)&Y_m(\beta r_i) &
I_m(\beta r_i) & K_m(\beta r_i) \\
J_m(\beta r_o) & Y_m(\beta r_o) & I_m(\beta
r_o) & K_m(\beta r_o) \\
J_{m-1}(\beta r_i) & Y_{m-1}(\beta r_i) & I_{m-1}(\beta r_i) &
-K_{m-1}(\beta
r_i) \\
J_{m-1}(\beta r_o) & Y_{m-1}(\beta r_o) &
I_{m-1}(\beta r_o) & -K_{m-1}(\beta r_o)
\end{array} \right] 
\end{eqnarray}
and $(1,B_{mn}, C_{mn}, D_{mn})$ is the eigenvector corresponding to
eigenvalue zero for each ${\beta_{mn}}$.  We tabulated the ${\beta_{mn}}$ at
each radius ratio $\alpha$ using a multiple root finding
routine.\cite{recipes} The determinant was found using an SVD algorithm, and
the eigenvectors by backsubstitution.\cite{recipes}  It was necessary to
rescale the third and fourth columns of ${\bf M}$ before taking the
determinant to avoid underflows and overflows.

The functions defined by Eq.~\ref{properstream} form a complete,
orthogonal
set with orthogonality condition
\begin{equation}
\int_{r_i}^{r_o} dr  ~r{\cal
C}_{m;n}{\cal C}_{m;p} = {\cal N}_{C_{m;n}}\delta_{np} \,.
\label{orthoC}
\end{equation}
The normalization ${\cal N}_{C_{m;n}}$ is given by
\begin{equation}
{\cal N}_{C_{m;n}} = {r_o}^2 {{\cal U}^2_{m;n}} (r_o) - {r_i}^2 {{\cal
U}^2_{m;n}}(r_i) \,,
\label{normC}
\end{equation}
where
\begin{equation}
{\cal U}_{m;n} (r) = J_m(\beta_{mn} r) +B_{mn} Y_m(\beta_{mn} r) \,.
\end{equation}
It is convenient to define a function ${\cal V}_{m;n}$ by
\begin{equation}
{\cal V}_{m;n} (r) = C_{mn}I_m(\beta_{mn} r) +D_{mn} K_m(\beta_{mn}
r) \,,
\end{equation}
so that ${\cal C}_{m;n}(r) = {\cal U}_{m;n} (r) + {\cal V}_{m;n} (r)$
and
\begin{equation}
\biggl(D_{*}D - \frac{m^2}{r^2} \biggr) C_{m;n} = -\beta_{mn}^2
({\cal U}_{m;n}-{\cal V}_{m;n}) \,. \label{eigeneq2}
\end{equation}

\subsection{Expansion functions for the potential}
\label{appendix_potential}
The potential perturbation satisfies homogeneous boundary
conditions at the inner and outer edges of the film and obeys
Eq.~\ref{NDCC_mode}, so that we can use an expansion in terms of the
eigenfunctions of the Laplacian operator,
\begin{equation}
\Biggl( D_{*}D - \frac{m^2}{r^2} \Biggr) \psi_{m;l} =
     -\chi_{ml}^2 \psi_{m;l} \,, \label{eigeneq3}
\end{equation}
which has solutions
\begin{equation}
\psi_{m;n}(r) = J_m(\chi_{mn} r) + b_{mn} Y_m(\chi_{mn} r) \,.
\label{properpot}
\end{equation}
The parameters $\chi_{mn}$ solve the secular equation $det~{\bf N}(\chi) =
0$, where
\begin{equation}
{\bf N}(\chi) = \left[ \begin{array}{cc}
J_m(\chi r_i) & Y_m(\chi r_i)\\
J_m(\chi r_o) & Y_m(\chi r_o)
\end{array}
\right] \,.
\end{equation}
and $(1,b_{mn})$ is the eigenvector corresponding to eigenvalue zero for
each ${\chi_{mn}}$. These constants were found by routines similar to the
ones used in Section~\ref{appendix_stream}. The functions defined by
Eq.~\ref{properpot} satisfy the orthogonality condition,
\begin{equation}
{\int_{r_i}}^{r_o} dr
{}~ r\psi_{m;n}\psi_{m;p} = {\cal N}_{\psi_{m;n}}\delta_{np} \,,
\label{orthopsi}
\end{equation}
where the normalization ${\cal N}_{\psi_{m;n}}$ is given by
\begin{eqnarray}
{\cal N}_{{\psi}_{m;n}} &=&\frac{1}{8}~~\Biggl[ {r_o}^2 \biggl(
\biggl[J_{m-1}(\chi_{mn} r_o) - J_{m+1}(\chi_{mn} r_o)\biggr] \nonumber \\
&+& b_{mn}\biggl[(Y_{m-1}(\chi_{mn} r_o) -
Y_{m+1}(\chi_{mn}r_o)\biggr]\biggr)^2 \nonumber \\
&-& {r_i}^2\biggl( \biggl[J_{m-1}( \chi_{mn} r_i) - J_{m+1}( \chi_{mn}
r_i)\biggr] \nonumber \\
&+& b_{mn}\biggl[Y_{m-1}( \chi_{mn} r_i) - Y_{m+1}( \chi_{mn}
r_i)\biggr]\biggr)^2\Biggr].
\end{eqnarray}


\begin{references}
\bibitem{smcfootnote} In Ref.\cite{smctheory}, the charge density is
eliminated using the Maxwell equation, so that the surface charges that we
consider here are not included.  The film is treated as if it were unbounded in the $z$ direction.
\bibitem{8cb} To control the nature of the ionic species in the
liquid crystal,
octylcyanobiphenyl (8CB), was doped with tetracyanoquinodimethane (TCNQ).
\bibitem{extrap_foot} We did not extrapolate our nonlocal annular
calculations to the limit of an infinitely dense relaxation grid, as was done
for the rectangular case in Ref.~\cite{linear}. This leads to a small difference between our numerical results for ${\cal R}^0_c$ in the limit $\alpha \rightarrow 1$ and those for the rectangular case, for which ${\cal R}_c = 91.84$.
\end{references}
\end{document}